\documentclass[manuscript]{aastex}
\usepackage{epsfig}                   
\usepackage{epstopdf}                   
\usepackage{multirow}
\usepackage{natbib}
\usepackage{txfonts}                  

\shorttitle{AGB stars at different metallicities (III)} \shortauthors{1,2,3,4,5}

\begin{document}

\newcommand{\ctan}{$^{13}$C($\alpha$,n)$^{16}$O }
\newcommand{\nean}{$^{22}$Ne($\alpha$,n)$^{25}$Mg }
\newcommand{\neag}{$^{22}$Ne($\alpha$,$\gamma$)$^{25}$Mg }
\newcommand{\ct}{$^{13}$C }
\newcommand{\nq}{$^{14}$N }
\newcommand{\odo}{$_\odot$}
\newcommand{\odos}{$_\odot$~}
\newcommand{\s}{{\it s}}

\title{Evolution, nucleosynthesis and yields
of AGB stars at different metallicities (III): intermediate mass
models, revised low mass models and the ph-FRUITY interface.}

  \author{S. Cristallo\altaffilmark{1,2}
    \affil{1 - INAF-Osservatorio Astronomico di Collurania, 64100 Teramo, Italy}\affil{2 - INFN Sezione Napoli, Napoli, Italy}}
  \and
  \author{O. Straniero\altaffilmark{1,3}
    \affil{1 - INAF-Osservatorio Astronomico di Collurania, 64100 Teramo, Italy}\affil{3 - INFN LNGS, Assergi, Italy}}
  \and
  \author{L. Piersanti\altaffilmark{1,2}
    \affil{1 - INAF-Osservatorio Astronomico di Collurania, 64100 Teramo, Italy}\affil{2 - INFN Sezione Napoli, Napoli, Italy}}
  \author{D. Gobrecht\altaffilmark{1}
    \affil{1 - INAF-Osservatorio Astronomico di Collurania, 64100 Teramo, Italy}}


\date{\today}

\begin{abstract}
We present a new set of models for intermediate mass AGB stars
(4.0, 5.0 and, 6.0 M$_\odot$) at different metallicities
(-2.15$\le$[Fe/H]$\le$+0.15). This integrates the existing set of
models for low mass AGB stars (1.3$\le$M/M$_\odot$$\le$3.0)
already included in the FRUITY database. We describe the physical
and chemical evolution of the computed models from the Main
Sequence up to the end of the AGB phase. Due to less efficient
third dredge up episodes, models with large core masses show
modest surface enhancements. The latter is due to the fact that
the interpulse phases are short and, then, Thermal Pulses are
weak. Moreover, the high temperature at the base of the convective
envelope prevents it to deeply penetrate the radiative underlying
layers. Depending on the initial stellar mass, the heavy elements
nucleosynthesis is dominated by different neutron sources. In
particular, the \s-process distributions of the more massive
models are dominated by the \nean~reaction, which is efficiently
activated during Thermal Pulses. At low metallicities, our models
undergo hot bottom burning and hot third dredge up. We compare our
theoretical final core masses to available white dwarf
observations. Moreover, we quantify the weight that intermediate
mass models have on the carbon stars luminosity function. Finally,
we present the upgrade of the FRUITY web interface, now also
including the physical quantities of the TP-AGB phase of all the
models included in the database (ph-FRUITY).
\end{abstract}

\keywords{Stars: AGB and post-AGB --- Physical data and processes:
Nuclear reactions, nucleosynthesis, abundances} \maketitle

\section{Introduction}\label{intro}

Asymptotic Giant Branch (AGB) stars are ideal laboratories to test
our understanding of stellar interiors. The evolution of those
objects is characterized by a sequence of burning and mixing
episodes, which carry the nuclear products synthesized in the
internal layers to the stellar surface. During the AGB, the
structure consists of a partially degenerate CO core, a He-shell
and a H-shell separated by a thin helium rich layer (the so-called
He-intershell) and, an expanded and cool convective envelope. The
surface luminosity is mainly sustained by the H-burning shell.
This situation is recurrently interrupted by the growing up of
thermonuclear runaways (Thermal Pulse, TP) in the He-intershell,
triggered by the activation of 3$\alpha$ reactions. The rate of
nuclear energy is too large to be carried away by radiation and,
thus, a convective shell develops, which makes the He-intershell
chemically homogenous. Then, the layers above expand and cool
until the convective shell quenches. If the expansion has been
large enough, the H-shell switches off and the convective envelope
can penetrate the H-exhausted and chemically enriched
He-intershell (this phenomenon is known as the Third Dredge Up,
TDU). Meanwhile, the products of internal nucleosynthesis appear
on the stellar surface. During the AGB, a strong stellar wind
erodes the convective envelope, thus polluting the interstellar
medium. AGB stars efficiently synthesize light (C, N, F and Na) as
well as heavy elements (those created via the slow neutron capture
process, the \s-process). The interested reader can find a vast
literature on AGB stars \citep{ir83,he05,stra06,kakka14}.\\
In order to properly describe the chemical evolution of the
hosting systems, sets of AGB yields as much uniform and complete
as possible are needed. In the past years, we made our AGB yields
available on the FRUITY database \citep{cri09,cri11}. Through a
web interface, we provide tables of isotopic and elemental
compositions as well as stellar yields of AGB stars. Up to date,
FRUITY includes low mass stars only (i.e. stars with initial
masses M$\le$ 3 M\odo). In this paper, we present a new set of AGB
models with larger initial masses (4$\le$ M/M\odo$\le$6). The
evolution of those objects resembles that of their low mass
counterparts, even if noticeable differences exist. In particular,
their larger initial masses produce more compact cores and, thus,
larger temperatures can be attained in their interiors. As a
consequence, the physical conditions to trigger the He-burning in
the He-intershell are attained earlier during the AGB phase with
respect to models with lower initial masses. Thus, the interpulse
phases are shorter, the TPs are weaker and the efficiency of TDU
is strongly reduced \citep{stra03}. Moreover, the larger the
initial stellar mass, the larger the mass extension of the
convective envelope (this fact implying a larger dilution of the
dredged up material). As a consequence, we obtain modest surface
chemical enrichments in the more massive AGBs. Furthermore, in
those objects processes like the Hot Bottom Burning (HBB;
\citealt{sugi71,iben73}) and the Hot-TDU (H-TDU;
\citealt{gosi2004,stra14}) can be active. During the HBB,
temperature becomes high enough to partially activate the CN cycle
at the base of the convective envelope. As a consequence,
considerable amounts of $^{13}$C and $^{14}$N can be produced. The
main effect of H-TDU, instead, is to limit the penetration of the
convective envelope itself, because the temperature for the
reactivation of the H-shell is attained soon. Thus, H-TDU further
weakens the TDU efficiency. \\
In AGB stars, two major neutron sources are at work: the \ctan and
the \nean reactions. In low mass stars, the dominant contribution
to \s-process nucleosynthesis comes from the \ctan reaction. The
\ct reservoir, the so-called \ct pocket, forms during TDU episodes
in the top layers of the H-exhausted He-intershell (for details
see \citealt{cri11}). In more massive AGBs, due to the limitations
of the H-TDU, the contribution from the \ct reaction is definitely
lower, while an important contribution comes from the
\nean\footnote{The abundant $^{22}$Ne is the final product of the
$^{14}$N($\alpha$,$\gamma$)$^{18}$F($\beta^+$)$^{18}$O($\alpha$,$\gamma$)$^{22}$Ne
nuclear chain.}. In fact, in those objects this reaction is
efficiently activated at the base of the convective shells
generated by TPs.  This neutron source significantly contributes
to the production of rubidium and light \s-process elements. Thus,
\s-process surface distributions with different shapes and
enhancements can be found by varying the metallicity and the
initial stellar mass. In fact, the three \s-process
peaks\footnote{The three \s-process components are: ls (Sr-Y-Zr),
hs (Ba-La-Nd-Sm) and lead (Pb)} receive different contributions
depending on the physical environmental conditions (radiative or
convective burning) and on
the neutron-to-seed ratio (which is related to the metallicity).\\
In this paper we also illustrate a new web interface (ph-FRUITY),
to access tables containing the evolution of the most relevant
physical quantities of our models.\\
This paper is structured as follows. In \S \ref{models} we
describe the main features of our stellar evolutionary code,
focusing on the most recent upgrades. In \S \ref{preagb} we
highlight the evolutionary phases prior to the AGB, which is
analyzed in \S \ref{agb}. In \S \ref{phf} we show the potentiality
of our new web ph-FRUITY interface. The nucleosynthesis of all
FRUITY models is discussed in detail in \S \ref{agbnuc}. Finally,
in \S \ref{conclu} we report the discussion and our conclusions.

\section{The models}\label{models}

As already outlined, models presented in this paper (4.0-5.0-6.0
M\odo) integrate the already existing set available on the FRUITY
database \citep{cri11}, currently hosting Low Mass Stars AGB
models (hereafter LMS-AGB; 1.3-1.5-2.0-2.5-3.0) with different
initial metallicities (-2.15$\le$[Fe/H]$\le$+0.15). We add a
further metallicity ($Z=0.002$, corresponding to [Fe/H]=-0.85) in
order to better sample the peak in the lead production (see
below). In Table \ref{tabini} we report all the models included in
the FRUITY database (in bold the models added with this work), by
specifying the initial He content, the [Fe/H] and the eventual
$\alpha$ enrichment. In the Table header we report both the [Fe/H]
and the corresponding total metallicity (which takes into account
for the eventual $\alpha$ enhancement). The isotopic initial
distribution of each model is assumed to be solar-scaled (apart
from eventual $\alpha$-enhanced isotopes, i.e. $^{16}$O,
$^{20}$Ne, $^{24}$Mg, $^{28}$Si, $^{32}$S, $^{36}$Ar and,
$^{40}$Ca). We adopt the solar distribution presented by
\cite{Lodders2003}. The models have been computed with the FUNS
evolutionary code \citep{stra06,cri09}. The physical evolution of
the star is coupled to a nuclear network including all isotopes
(from hydrogen to lead, at the termination of the \s-process
path). Thus, we do not need to perform any post-process
calculation to determine the nucleosynthetic yields. The list of
reactions and the adopted rates are the same as in \cite{cri11}.
Among the various physical processes, convection and mass-loss
mainly affect the AGB evolution (and, thus, the correlated stellar
yields and surface distributions). We determine convective
velocities following the prescriptions of the Mixing Length Theory
(MLT; \citealt{boom}), according to the derivation of
\cite{cox68}. In the framework of the MLT, in correspondence to a
convective border the velocity is zero, if the adiabatic
temperature gradient presents a smooth profile\footnote{We remind
that the velocity is proportional to the difference between the
radiative and the adiabatic temperature gradients.}. However,
during a TDU episode there is a sharp discontinuity in the opacity
profile (and, thus, in the radiative gradient). This makes the
convective/radiative interface unstable. In order to handle such a
situation, we apply an exponentially decreasing profile of
convective velocities below the formal Schwarzschild border
\citep{stra06}. This implies a more efficient TDU and, as a
by-product, the formation of a tiny $^{13}$C pocket. In fact, such
a non-convective mixing allows some protons to penetrate the
formal border of the convective envelope. Those protons are
captured by the abundant $^{12}$C (the product of 3$\alpha$
processes) leading to the formation of a region enriched in
$^{13}$C (commonly known as the $^{13}$C-pocket). In \cite{cri09}
we demonstrated that the extension in mass of the \ct pocket
decreases along the AGB (thus with increasing core masses),
following the shrinking and the compression of the He-intershell
region. Therefore, we expect that the contribution to the
\s-process nucleosynthesis from the \ct pocket is strongly reduced
in massive AGBs with respect to their low-mass counterparts (see
\S \ref{agb}). In the following, we define Intermediate Mass Stars
(IMS) those approaching the TP-AGB phase
with a mass of the H-exhausted core greater than 0.8 M\odos (see \S \ref{preagb}).\\
Another very uncertain physical input for AGB models is the
mass-loss rate, which largely determine, for instance, the
duration of the AGB and the amount of H-depleted dredged-up
material after each TDU. During the AGB, large amplitude
pulsations induce the formation of shocks in the most external
stellar layers. As a result, the local temperature and density
increase and a rich and complex chemistry develops, leading to the
creation of molecules and dust grains. Those small particles
interact with the radiation flux and drive strong stellar winds.
Available observational data indicate that in galactic AGB stars
the mass loss ranges between $10^{-8}$ and $10^{-4}$
$M_{\odot}$/yr, with a clear correlation with the pulsational
period \citep{vw93}, at least for long periods (see
\citealt{utte13}). Since the latter depends on the variations of
radius, luminosity and mass, a relation between the mass loss rate
and the basic stellar parameters can be derived. By adopting a
procedure similar to \cite{vw93}, we revised the mass loss-period
relation, taking in to account more recent infrared observations
of AGB stars (see \citealt{stra06} and references therein) and
basing on the observed correlation between periods and
luminosities in the K band (see e.g. \citealt{white03}). The few
pulsational masses derived to date for AGB stars \citep{wood07} do
not allow the identification of trends in the mass loss-period
relation as a function of the stellar mass\footnote{Note that
\cite{vw93} delayed the onset of the super-wind phase in stars
with masses greater than 2.5 M\odo.}. Thus, we apply the same
theoretical recipe for the whole mass range in our models, even if
other mass-loss prescriptions are available for luminous oxygen
rich AGB giants. For instance, we could use the mass-loss rate
proposed by \citet{vl05}. However, when applying that formula to
our low metallicity models, the mass loss practically vanishes
and, therefore, we exclude it.\\
When dealing with C-rich objects, particular attention must be
paid to the opacity treatment of the most external (and cool)
regions. As already discussed, molecules efficiently form in those
layers. Depending on the C/O ratio, O-bearing or C-bearing
molecules form, leading to dichotomic behaviors in the opacity
regime. C-bearing molecules, in fact, are more opaque and, thus,
increase the opacity of the layer where they form. As a
consequence, the radiation struggles to escape from the stellar
structure and, as a consequence, the most external layers expand
and cool. This naturally implies an enhancement in the mass-loss
rate, which strongly depends on the stellar surface temperature.
We demonstrated that the use of low temperature C-bearing
opacities has dramatic consequences of the physical evolution of
AGB stars (\citealt{cri07}; see also \citealt{marigo2002} and
\citealt{vema}). For solar-scaled metallicities, we adopt
opacities from \cite{lear}, while for $\alpha$ enhanced mixtures
we use the AESOPUS tool \citep{esopus}, which allows to freely
vary the chemical composition. In calculating the IMS-AGB models,
we found an erroneous treatment of opacities in the most external
layers of the stars, enclosing about 2\% of the total mass. Then,
we verified one by one all the low mass models already included in
the FRUITY database and for some of them we found significative
variations in the final surface composition. We discuss this
problem in \S \ref{agbnuc}.

\section{From the pre-main sequence to the thermal pulse AGB phase}\label{preagb}

\begin{figure*}[tpb]
\centering
\includegraphics[width=\textwidth]{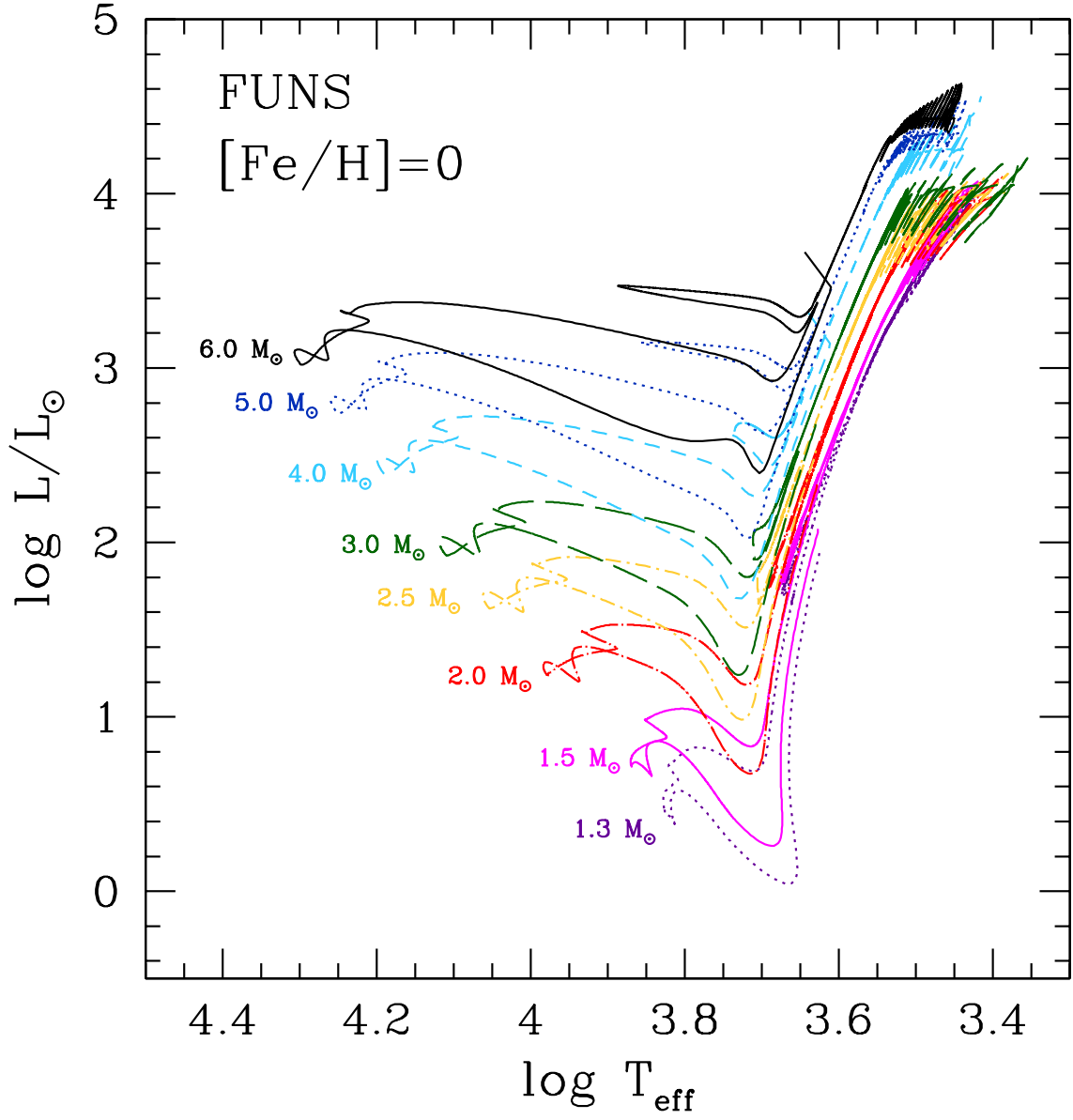}
\caption{Hertzsprung-Russell diagram for models with initial solar
metallicity. } \label{fig1}
\end{figure*}

\begin{figure*}[tpb]
\centering
\includegraphics[width=\textwidth]{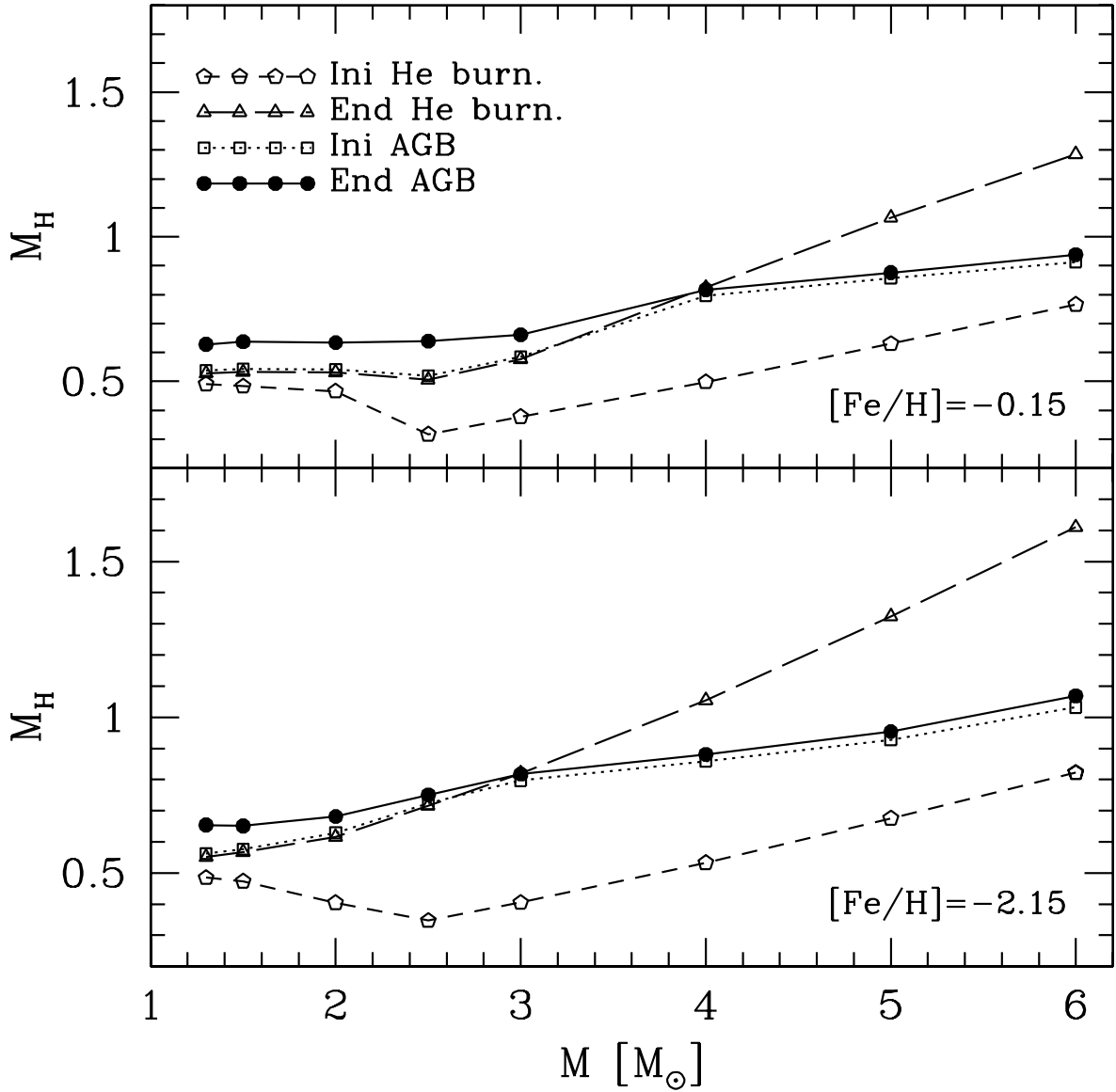}
\caption{He-exhausted core masses at the beginning of central He
burning (pentagons), at the end of central He burning (triangles),
at the beginning of the AGB (squares) and, at the end of AGB
(circles). } \label{fig2}
\end{figure*}

We follow the evolution of the models listed in Table \ref{tabini}
from the pre-Main Sequence up to the AGB tip. The computations
terminate when the H-rich envelope is reduced below the threshold
for the TDU occurrence. The Hertzsprung-Russell tracks of the
solar metallicity set
are shown in  Figure \ref{fig1}.\\
In this Section we briefly revise the evolution until the
beginning of the TP-AGB phase. For a more detailed description of
these phases see \cite{dom99}. All the evolutionary sequences
start from a homogeneous and relatively cool model relaxed on the
Hayashi track, i.e., the first fully convective models in
hydrostatic and thermal equilibrium. As usual, stars enter the
Main Sequence (MS) when all the secondary isotopes involved in the
p-p chain and the CNO cycle attain the equilibrium in the central
region. Table \ref{tabms} report MS lifetimes for the whole data
set\footnote{ For completeness, we also include data already
reported in \citet{cri11}.}. In Equation 1, we provide a simple
interpolation formula linking the MS lifetime to the initial mass
and metallicity of the model:
\begin{equation}
\tau_{\rm MS}=9.775+9.898*Z-1.460*ln(M)+0.152*(ln(M))^2
\end{equation}\label{fit}
The variations of these lifetimes reflect the well-known
non-linearity of the mass-luminosity relation for MS stars. The
less massive models (1.3 M$_\odot$) mark the transition between
the lower MS (consisting of stars whose luminosity is mainly
controlled by the p-p nuclear chain and characterized by a
radiative core and a convective envelope) and the upper MS (whose
stars burn H through the CNO cycle and develop a convective core,
while their envelope remains fully radiative). The convective core
attains a maximum extension just after the beginning of the MS.
Then, its extension decreases, as H is converted into He and,
consequently, the radiative opacity decreases. The maximum
extension of the convective core is reported in Table
\ref{tabcch}. No convective core overshoot has been assumed in
these models. Central convection eventually disappears when the
central H mass fraction drops below $\sim 0.1$. Then, an overall
contraction occurs. The tip of the MS, i.e., the relative  maximum
in the luminosity, is attained when the central H goes to $\sim
0$. Then, before He ignition, all the models, except the more
massive with $Z\le 10^{-3}$, experience a deep mixing episode, the
so-called first dredge up (FDU).
The following development of the He burning depends on the
equation of state of the He-rich core. The values of the He-core
masses at the ignition well represent this occurrence, as reported
in Table \ref{tabinihe}. For the more massive models (M$\ge 3$
M$_\odot$), they are essentially determined by the extension of
the convective core during the main sequence. After the core-H
burning, these stars rapidly proceed toward a quiescent He
ignition that occurs at relatively low density. On the contrary,
for less massive stars the core mass slowly grows during the RGB,
because of the shell H-burning.  For this reason, in stars with
M$< 2$ M$_\odot$, the central density grows up to $10^5$-$10^6$
g/cm$^3$ so that the pressure is mainly controlled by degenerate
electrons. Under these conditions, the He ignition proceeds
through a thermonuclear runaway (He flash), when the core mass
exceeds a critical value of about 0.5 M$_\odot$. In slightly more
massive objects ($2<M/M_\odot<$3), the electron degeneracy is
weaker and the critical core mass for the He ignition is reduced
down to $\sim0.3$ M$_\odot$ (see \citealt{pd09}). This behavior is
also illustrated in Figure \ref{fig2}; the minimum core mass at
the He ignition is found for stellar masses $2.0< $M/M$_\odot <
2.5$ (see also \citealt{swe90,dom99}).
Table \ref{tabheb} reports the core He burning lifetimes. The
variation of this quantity reflects the variations of the core
mass at the He ignition. In fact, the lower the core mass, the
fainter the He burning phase and, in turn, the longer the
lifetime. As a result, the longest core He burning lifetimes are
attained for stellar masses between 2 and 2.5 M$_\odot$. Our core
He burning models include specific treatments of the instability
occurring at the external border of the convective core and
semi-convection, as described in \citet{stra03wd}. During this
phase, the core mass further increases, due to the work done by
the shell H burning (see Figure \ref{fig2}). For high
metallicities, at the end of the core He burning phase, the core
mass is nearly constant for M$<3$ M$_\odot$ and rapidly increases
for larger stellar masses. This limit is smaller at lower Z. The
masses of the H-exhausted core at the end
of the central He burning phase are reported in Table \ref{tabheend}.\\
During the early-AGB phase, an He burning shell forms and advances
in mass at a rate much higher than that of the pre-existing H
burning shell. In the more massive models, the H burning dies down
and another mixing episode occurs (the Second Dredge Up, SDU),
owing to the expansion powered by the He burning and the
consequent cooling of the envelope. The lowest stellar mass
undergoing a SDU is the 3 M$_\odot$ at $Z$=0.0001 and the 4
M$_\odot$ at $Z$=0.02. Table \ref{tabiniagb} lists the core mass
at the onset of the first thermal pulse. It practically coincides
with the value attained at the end of the core-He burning, except
for stars undergoing the SDU,
as clearly shown in Figure \ref{fig2}.\\
Due to the mass lost in the previous evolutionary phases, stars
attain the AGB with masses lower than the initial ones. In our
models, we adopt a Reimers' parametrization of the mass-loss rate
(with $\eta=0.4$) up to the first TP. In general, only stars with
M$\le 2$ M$_\odot$ (those developing a degenerate He-rich core
during the RGB) lose a non negligible fraction of
their initial mass (see Table \ref{tabmtotiniagb}).\\
The modifications of the chemical compositions induced by the FDU
and (eventually) the SDU are reported in Tables \ref{tabfdusdu}
and \ref{tabfdusdu2} for two different metallicities ($Z=1\times
10^{-2}$ and $Z=2.4\times 10^{-4}$, respectively). As expected,
after a dredge up episode (FDU or SDU) the models show an increase
in the surface helium abundance as well as modified CNO isotopic
ratios. It should be noted that the abundances observed at the
surface of low mass (M$<$ 2.0 M$_\odot$) giant stars at the tip of
the RGB often differ from those at FDU due to the presence of a
non-convective mixing episode, which links the surface to the hot
layers above the H-burning shell. This occurs when stars populate
the so-called bump of the luminosity function (see
\citealt{palmerini2011} for a discussion on the various proposed
physical mechanisms triggering such a mixing; see also
\citealt{nb14}). Those chemical anomalies have been observed, for
instance, in low metallicity stars \citep{gratton00} and measured
in oxide grains (Al$_2$O$_3$) of group 1 \citep{nittler97}.
Considering that, up-to-date, no definitive theoretical recipe
exists for this non-convective mixing, our models do not include
any RGB extra-mixing. Among the isotopes reported in Table
\ref{tabfdusdu} and Table \ref{tabfdusdu2}, the most sensitive
isotopes to an extra-mixing process should be $^{13}$C and
$^{18}$O. This has to be kept in mind when adopting our isotopic
abundances.

\section{The TP-AGB phase (I): physics }\label{agb}

\begin{figure*}[tpb]
\centering
\includegraphics[width=\textwidth]{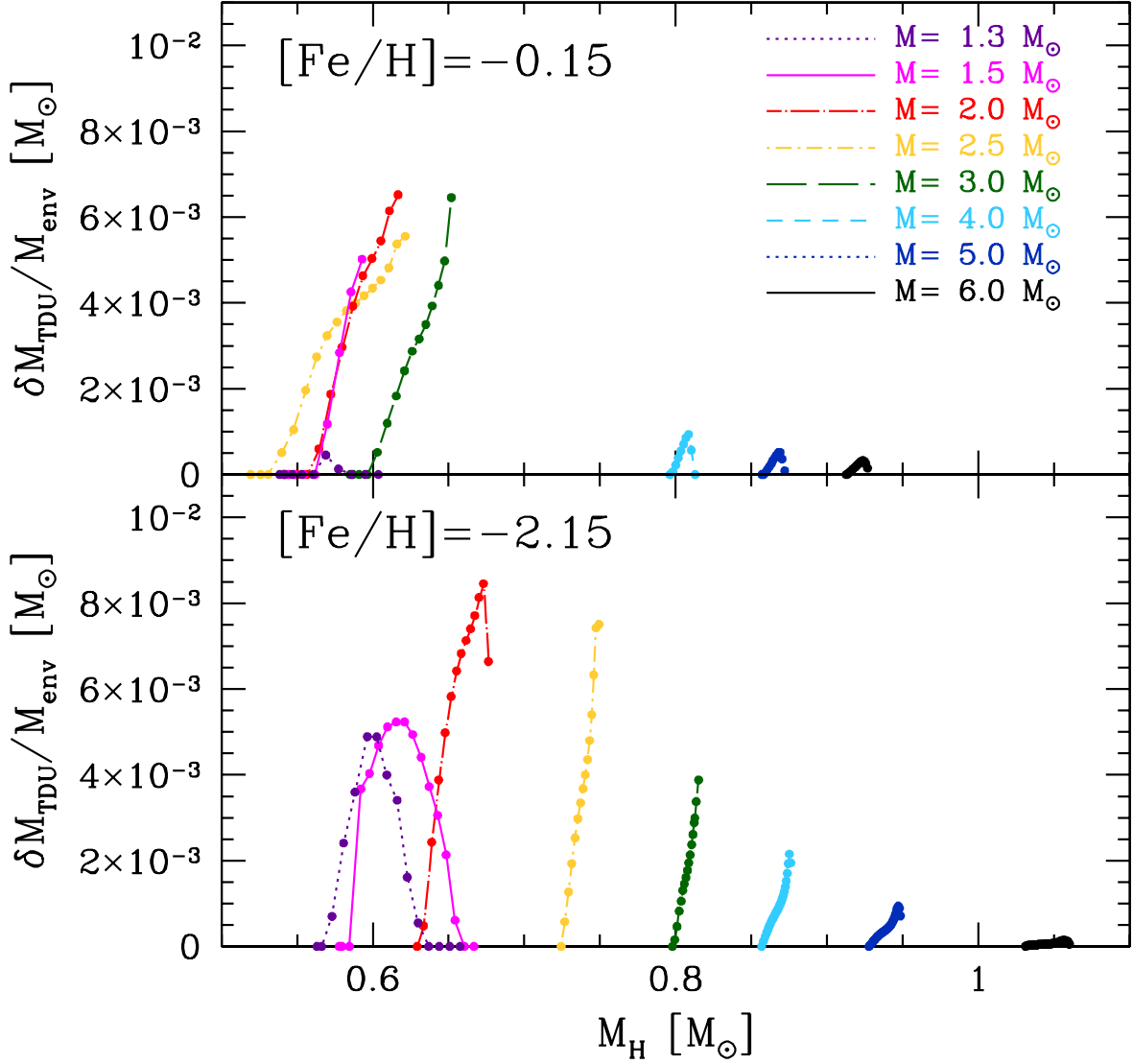}
\caption{Ratio between the mass of H-depleted dredged-up material
($\delta$M$_{\rm TDU}$) and the envelope mass (M$_{\rm env}$) for
different masses at Z=$10^{-2}$ (upper panel) and Z=$2.4\times
10^{-4}$ (lower panel). } \label{fig3}
\end{figure*}

FUNS models with mass 1.3$\le$M/M\odo$\le$3.0 has been extensively
analyzed in \cite{cri09} and \cite{cri11}. However, in order to
provide a general picture of stellar evolution during the AGB
phase, some of their physical properties will be addressed
here again. \\
In order to evaluate the behavior of the TDU mechanism as a
function of the mass and the metallicity, we plot the ratio
between the mass of H-depleted dredged-up material at each TDU
($\delta M_{TDU}$) and the envelope mass as a function of the mass
of the H-exhausted core in Figure \ref{fig3}. Such a quantity
provides an estimation of the TDU efficiency in polluting the
convective envelope. This Figure shows that at Z=$10^{-2}$ (upper
panel) a star with initial mass M=1.3 M\odos is close to the lower
mass limit to experience TDU. The maximum TDU efficiency is
reached for the 3 M\odos model. Then, there is an abrupt drop in
the TDU efficiency in correspondence to the 4 M\odos model. In
fact, the physical structure of this model is deeply different
with respect to those of lower masses. In particular, the mass of
the H-exhausted core (M$_H$) is definitely larger.
\begin{figure*}[tpb]
\centering
\includegraphics[width=\textwidth]{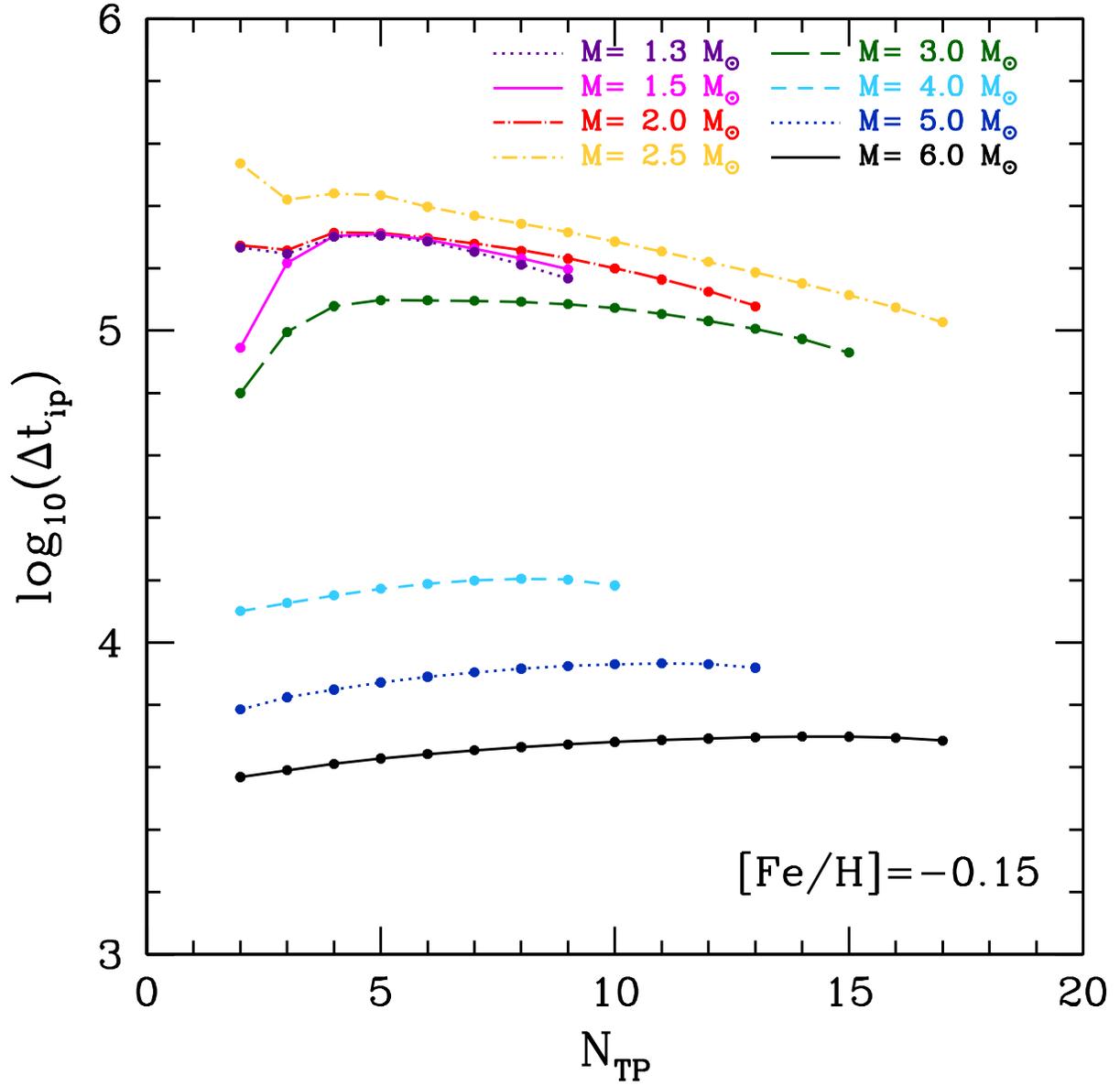}
\caption{Interpulse duration for models with Z=10$^{-2}$.}
\label{fig4}
\end{figure*}
\begin{figure*}[tpb]
\centering
\includegraphics[width=\textwidth]{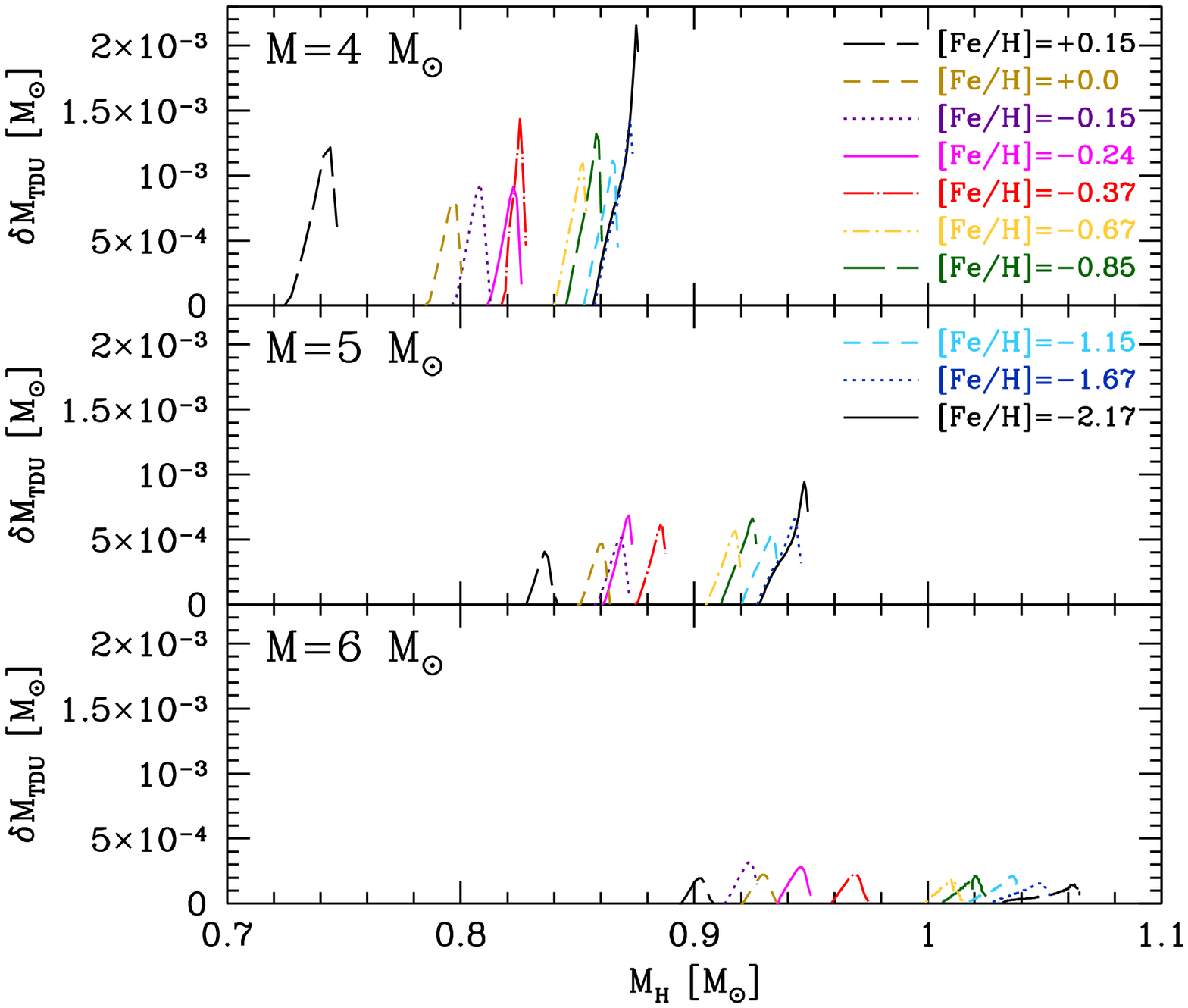}
\caption{Mass of H-depleted dredged-up material at each TDU
for models with initial mass M=4 M\odos (upper panel), M=5 M\odos
(middle panel) and M=6 M\odos (lower panel). } \label{fig5}
\end{figure*}
This implies a larger compression of the H-exhausted layers and,
thus, the He-intershell is thinner and hotter. As a consequence,
the time needed to reach the ignition conditions for the 3$\alpha$
process during H-shell burning is shorter. Hence, the interpulse
period decreases and, finally, the TDU efficiency is lower
\citep{stra03}. In Figure \ref{fig4} we report the interpulse
phase duration ($\Delta t_{\rm ip}$) for models with Z=$10^{-2}$.
While for models with M$\le$ 3 M\odos the $\Delta t_{\rm ip} \ge
10^5$, for larger masses it decreases to $10^4$ yr and 5000 yr for
the 4 M\odos and 6 M\odo, respectively. As shown in the lower
panel of Figure \ref{fig3}, at low metallicities even the lowest
masses (1.3-1.5 M\odo) experience a deep TDU, due to the low CNO
elemental abundances in the envelope (which implies a reduced
H-shell efficiency). Moreover, the transition between LMS-AGBs and
IMS-AGBs is smoother, as the 2.5 and 3.0 M\odos models start the
TP-AGB phase with definitely larger core masses with respect to
their high-metallicity counterparts. More massive models (4-5-6
M\odo) are characterized by a very low TDU efficiency (as their
metal-rich counterparts), but show a definitely larger number of
TPs (see Table \ref{tabtdu}). This is due to the fact that the
stellar structure is more compact and, thus, the external layers
are hotter. As a consequence, the mass-loss erodes the convective
envelope at a lower rate and the star experiences a larger number
of TPs. In Figure \ref{fig5} we report the $\delta M_{\rm TDU}$
for all the computed IMS-AGB models. As already highlighted, in
the 6 M\odos model there is no trend with the initial metallicity,
the TDU efficiency being always very low. In the 5 M\odos models
there is a slight increase of TDU efficiency at low metallicities.
For $Z \le Z_\odot$, the 4 M\odos models show a clear increase of
the TDU efficiency by decreasing the initial iron content. The
$Z=2\times 10^{-2}$ model represents an exception, because it
shows a net increase of the TDU efficiency. In fact, the core at
the beginning of the TP-AGB is less massive than that of models
with lower $Z$ and, thus, an increase of the TDU efficiency is
expected. In Table \ref{tab_dmtdu} we report the cumulative
dredged up mass (in M$_\odot$) for different initial masses and
metallicities ($\Delta M_{\rm TDU}$). As expected, models with the
largest $\Delta M_{\rm TDU}$ are in the range 2-2.5 M\odos and,
thus, the major pollution of the interstellar medium is expected
from these objects (see \S \ref{agbnuc}). Another way to evaluate
the TDU efficiency is to analyze the behavior of the $\lambda$
values, defined as the ratios between the mass of H-depleted
dredged-up material and the mass growth of the H-exhausted core
during the previous interpulse phase (see Figures \ref{fig6} and
\ref{fig7}). At solar-like metallicities, the maximum values of
$\lambda$ we obtain is $\sim 0.5$, for stars with M$\ge 2.0 $
M\odo. At low Z,
instead, the $\lambda$ grows up to 0.8, implying a larger TDU efficiency.\\
\begin{figure*}[tpb]
\centering
\includegraphics[width=\textwidth]{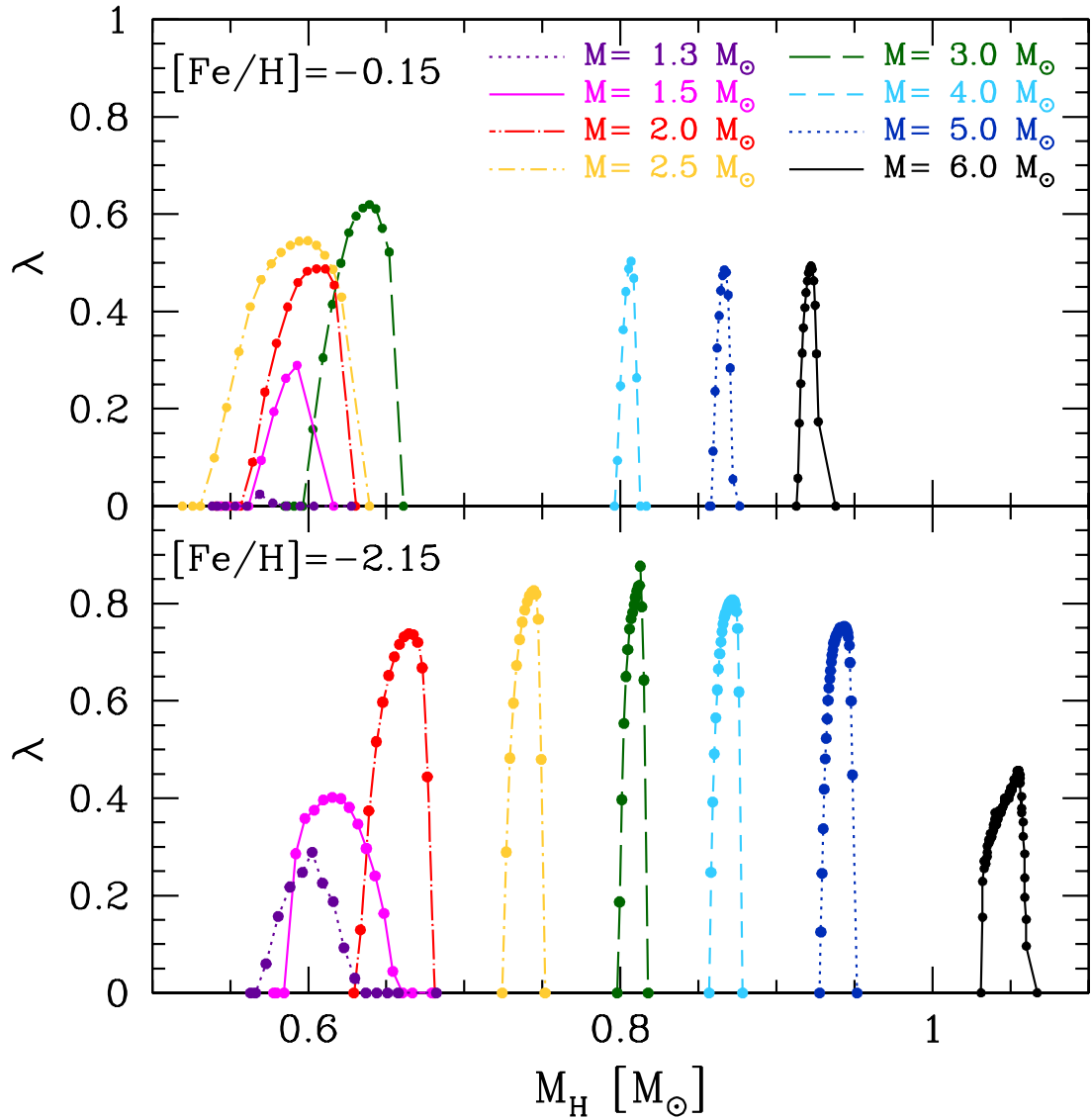}
\caption{As in Figure \ref{fig3}, but for the $\lambda$ values,
defined as the ratios between mass of H-depleted dredged-up
material and the growth of the H-exhausted core during the
previous interpulse phase. } \label{fig6}
\end{figure*}
\begin{figure*}[tpb]
\centering
\includegraphics[width=\textwidth]{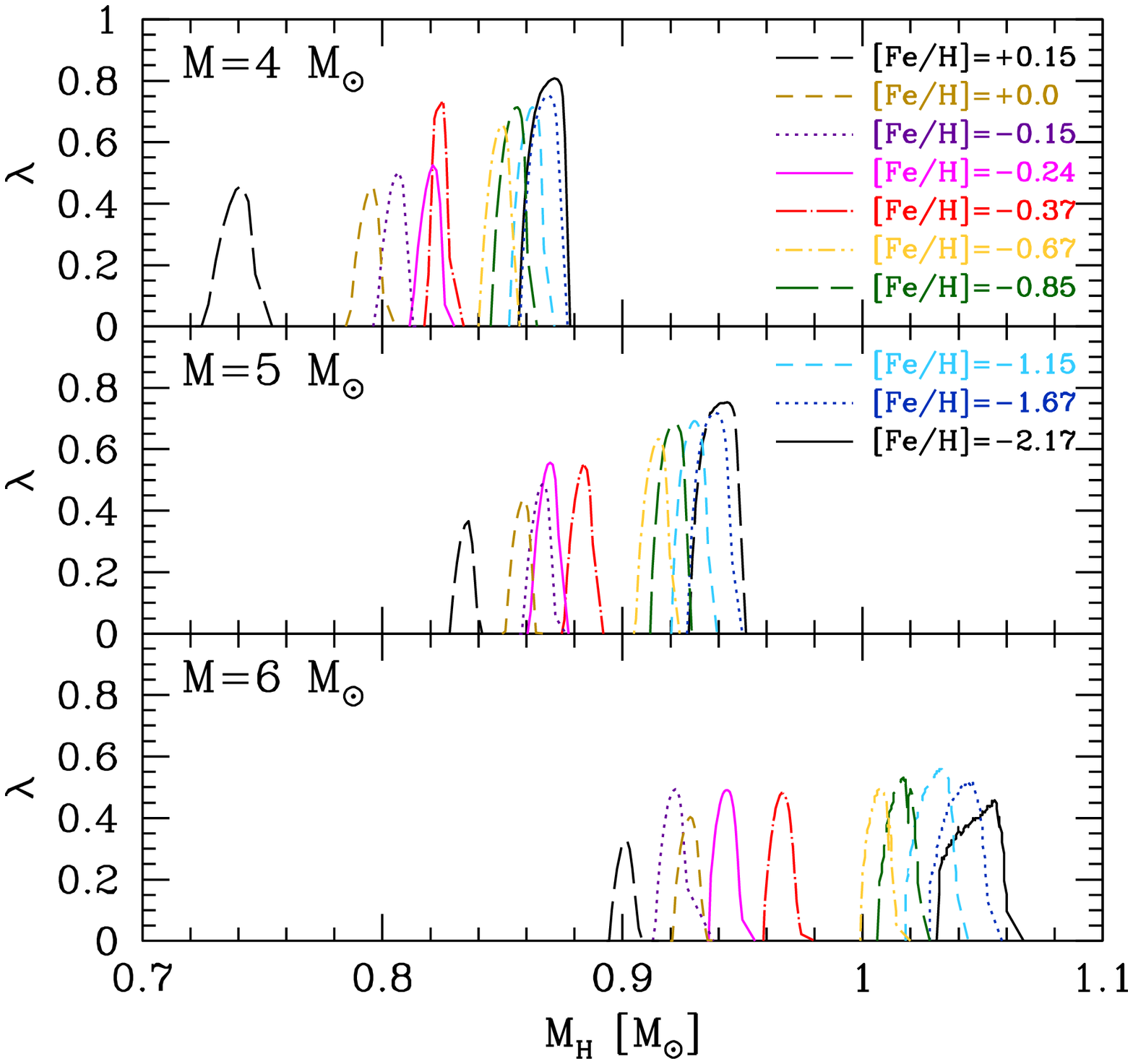}
\caption{As in Figure \ref{fig5}, but for the $\lambda$ values.} \label{fig7}
\end{figure*}
Depending on the final C/O ratio, stars are classified as C-rich
objects (C/O$>$1) or O-rich objects (C/O$<$1). The final surface
C/O ratio depends on many factors, whose effects are not easy to
be disentangled. Among them, the TDU efficiency and the mass-loss
rate play a major role. The uncertainties affecting these
phenomena have been extensively reviewed in
\cite{ventura2005_1,ventura2005_2}. They show that very different
results can be obtained by modifying, within uncertainties, the
recipe adopted to treat them. In \S \ref{conclu} we will compare
our models to similar computations described in the recent
literature.\\
While the TDU efficiency is strictly connected to the mixing
algorithms adopted to compute the models (mixing scheme; treatment
of convective borders; etc), the number of TDU mainly depends on
the mass-loss mechanism. The latter, in fact, erodes the mass of
the convective envelope and determines the dilution factor between
the cumulatively dredge up material and the envelope mass itself.
Thus, if the envelope mass is not too large and the number of
experienced TDUs is high enough, the model shows C/O$>$1 at the
surface. Once again, the duration of the C-rich phase depends on
the efficiency of mass-loss in eroding the convective envelope. As
already stressed, the presence of carbon bearing molecules locally
increases the opacity. Thus, we expect an increase of the mass
loss rate when
the C/O becomes greater than 1. \\
In Table \ref{tab_agb}, we report the total TP-AGB lifetimes
($\tau_{\rm TP-AGB}$) for all FRUITY models. Models which are
still O-rich at the end of their evolution are labelled as
\textbf{O}, while models ending their evolution with a surface
C/O$>$1 are labelled as \textbf{C}. Numbers in brackets refer to
the percentage of the TP-AGB phase spent in the C-rich regime. As
it can be seen, all the models become C-rich at low metallicities
and spend the majority of their TP-AGB lifetime in the C-rich
regime. This can also be appreciated in Figure \ref{fig8}, where
we report $\tau_{\rm TP-AGB}$ (histograms) and the corresponding
time spent during the C-rich regime (shaded histograms) for three
different metallicities ($Z=2\times 10^{-2}$: upper panel;
$Z=6\times 10^{-3}$: middle panel; $Z=1\times 10^{-3}$: lower
panel). In general, the larger the metallicity, the longer the
TP-AGB lifetimes and the lower the time fraction spent in the
C-rich regime. The lifetimes of IMS-AGBs are definitely shorter
(about a factor 10) with respect to LMS-AGBs. From Figure
\ref{fig8}, it turns out that stars with the longest TP-AGB
lifetimes have M=2.5 M\odos for large and intermediate
metallicities. At low Z, instead, we find a monotonic trend,
with the lowest masses showing the longest $\tau_{\rm TP-AGB}$.\\
\begin{figure*}[tpb]
\centering
\includegraphics[width=\textwidth]{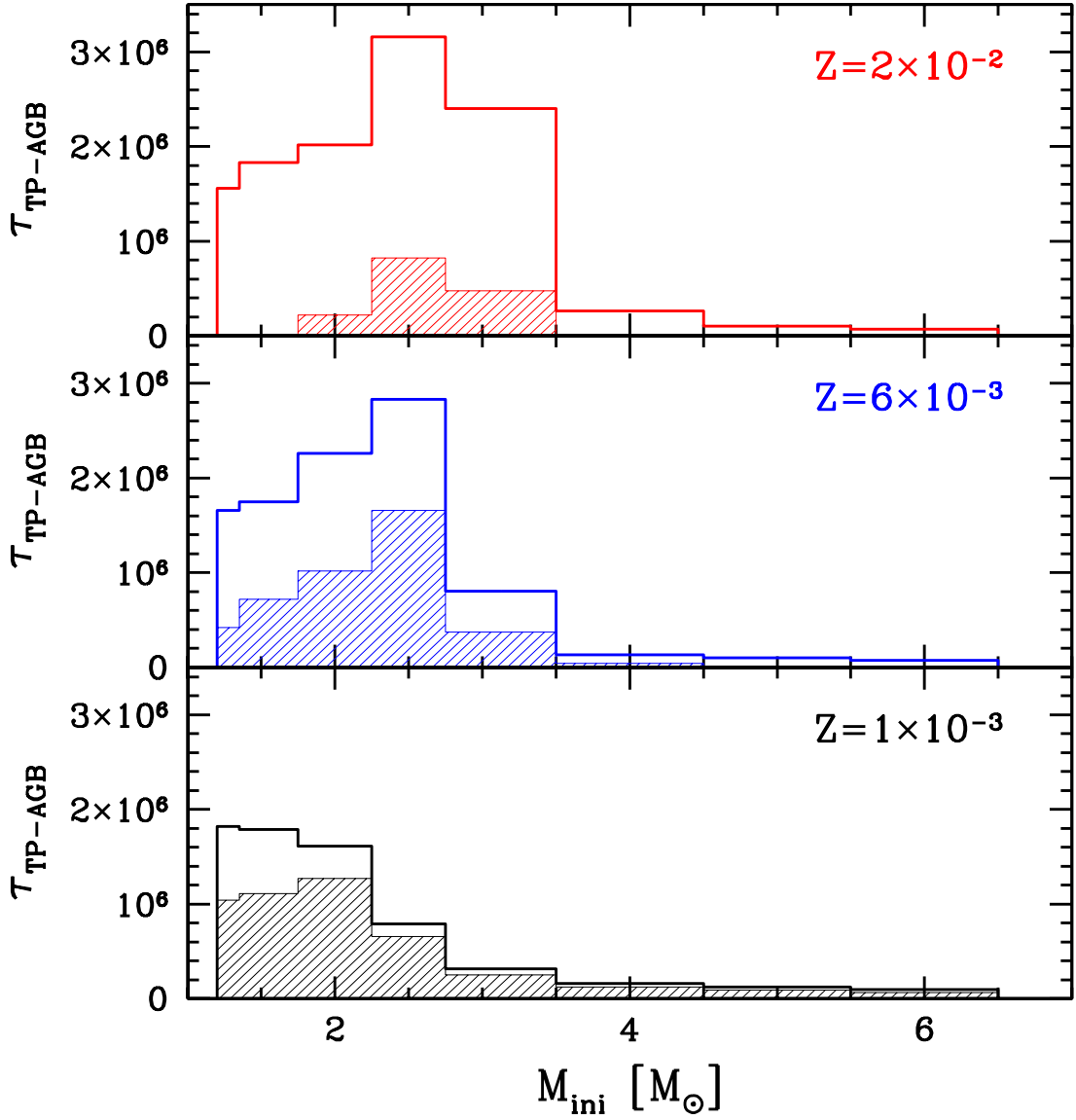}
\caption{TP-AGB lifetimes for three selected metallicities. Shaded
histograms refer to the C-rich phase of the models.} \label{fig8}
\end{figure*}
\begin{figure*}[tpb]
\centering
\includegraphics[width=\textwidth]{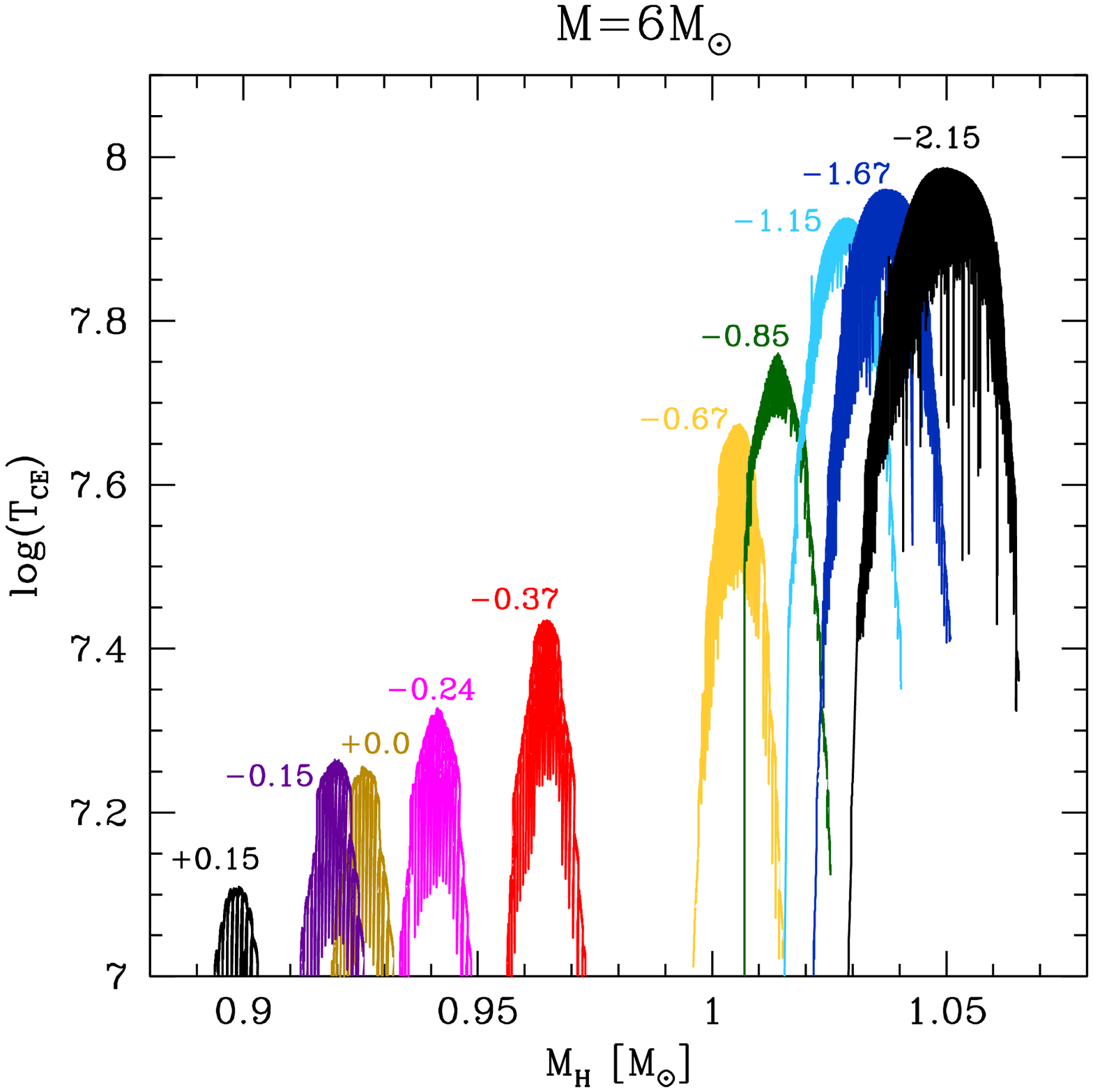}
\caption{Evolution of the temperature attained at the base of the
convective envelope during the TP-AGB phase of 6 M\odos models at
different [Fe/H]. } \label{fig9}
\end{figure*}
Note that the HBB and the H-TDU could be at work during the TP-AGB
phase. Both phenomena are able to modify the surface C/O ratio. In
Figure \ref{fig9}, we report the maximum temperature attained at
the base of the convective envelope during the TP-AGB phase of 6
M\odos models at different metallicities. In order to efficiently
activate the HBB, the base of the convective envelope should
attain 80 MK. In our models, this condition is fulfilled at the
lowest metallicities for the largest masses (5-6 M\odo) only.
Another interesting phenomenon, possibly working during the TP-AGB
phase, is the H-TDU. In this case, the temperature at the base of
the convective envelope is high enough to restart the H-burning
during a TDU episode. As a result, protons are mixed and burnt
on-flight. In our models, H-TDU is activated at low metallicities
only, with particular high efficiencies in the more massive
models. The effects of HBB and H-TDU on the nucleosynthesis of our
models are discussed in \S \ref{agbnuc}. Considering that the
envelopes of AGB stars are very expanded ($R\sim 200-1000$
R$_\odot$) and that the average convective velocity is low ($v\sim
10^5$ cm s$^{-1}$), for some isotopes the convective turnover
timescale is longer than the proton capture timescales at the base
of the convective envelope. In order to properly treat those
processes, the computation of the chemical evolution should be
performed by coupling mixing and burning. We intend to address it
in a future study. Models presented in this paper have been
calculated with a 3 step process. First, we burn chemicals over a
model time step. Then, we mix them in convective regions following
a time-dependent mixing scheme derived from an algorithm proposed
by \cite{sp80}\footnote{We assume that neutrons are at the local
equilibrium and, hence, they are not mixed.}. Finally, we burn
again chemicals in convective regions only for a fraction
(10$^{-2}$) of the model time step. This is done in order to allow
isotopes to reach their equilibrium abundance in the case of their
burning timescale being lower
than the convective turnover timescale.\\
\begin{figure*}[tpb]
\centering
\includegraphics[width=\textwidth]{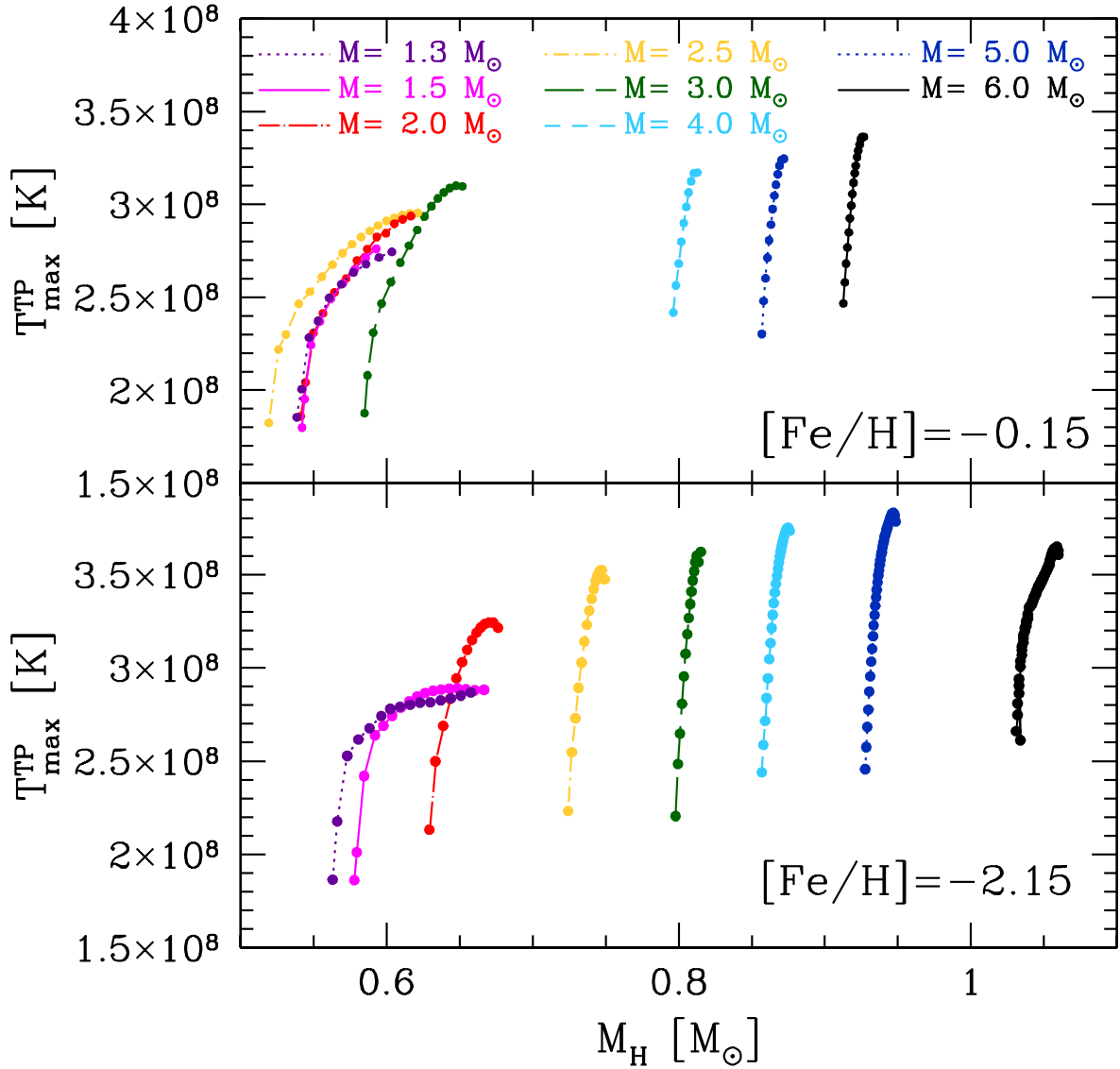}
\caption{As in Figure \ref{fig3}, but for the maximum temperature
attained at the bottom of the convective zone generated by a TP. }
\label{fig10}
\end{figure*}
\begin{figure*}[tpb]
\centering
\includegraphics[width=\textwidth]{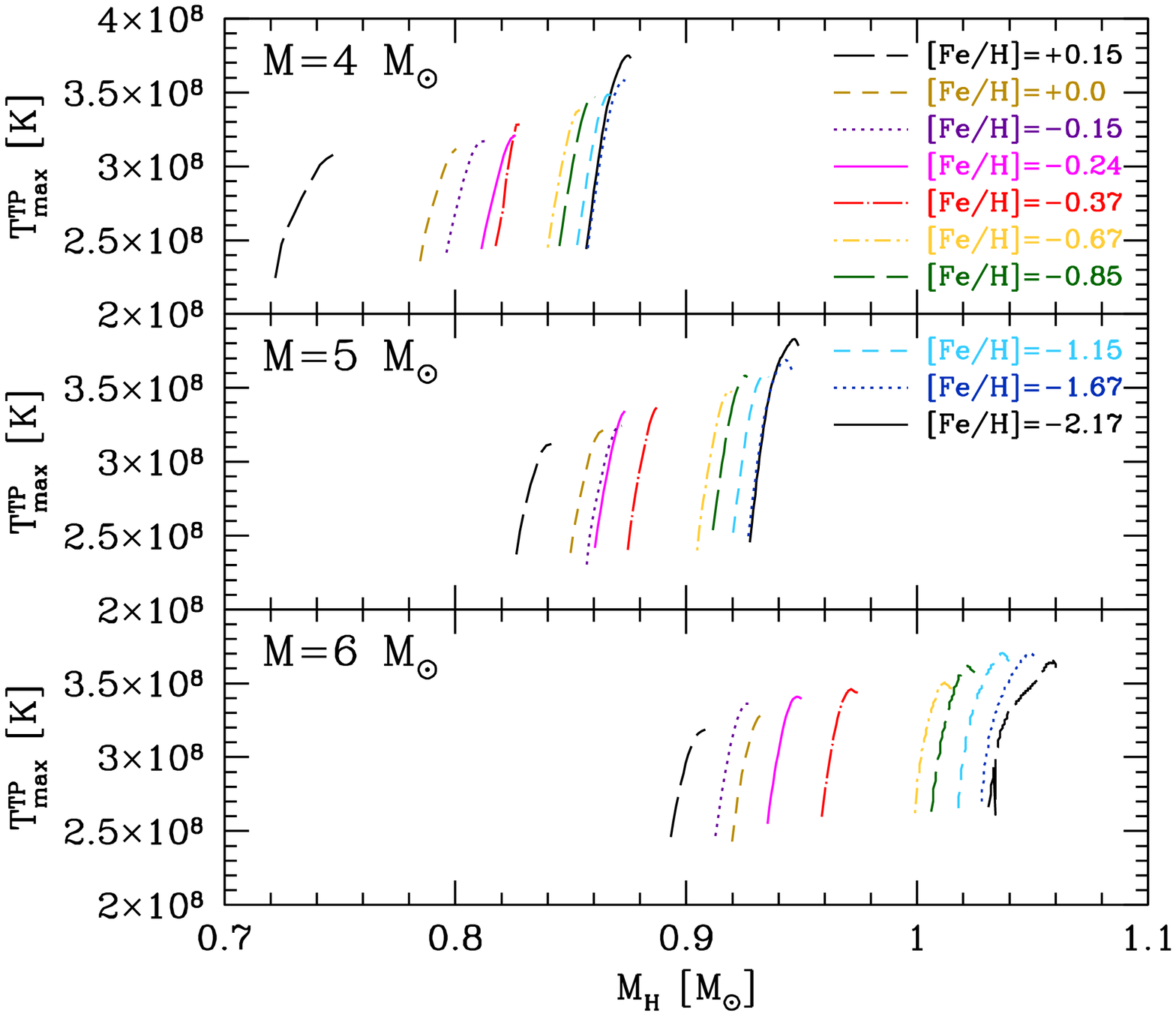}
\caption{As in Figure \ref{fig5}, but for the maximum temperature
attained at the bottom of the convective zone generated by a TP. }
\label{fig11}
\end{figure*}
Another interesting feature of IMS-AGB models is the large
temperature reached at the bottom of the convective shells
generated by TPs (T$^{\rm TP}_{\rm max}$). Depending on T$^{\rm
TP}_{\rm max}$, the \nean reaction can be efficiently activated.
This can lead to a second neutron burst (additional to that from
the \ctan reaction), with important consequences on the \s-process
nucleosynthesis (see \S \ref{agbnuc}). From an inspection of
Figures \ref{fig10} and \ref{fig11}, it can be noticed that,
during the AGB, T$^{\rm TP}_{\rm max}$ progressively increases,
reaches a maximum, and then slightly decreases. This quantity
depends on the core mass, the envelope mass and the metallicity
(see \citealt{stra03}). As T$^{\rm TP}_{\rm max}$ scales with the
core mass, we expect the largest temperatures to be attained in
the models with the largest initial masses. This is clearly shown
in Figure \ref{fig10}, where it can be assessed that low mass
stars barely reach T$^{\rm TP}_{\rm max} \sim 3\times 10^8$ K,
while IMS-AGB easily go beyond this limit. The dependence on the
initial metallicity is also evident, the models with low Z showing
definitely larger T$^{\rm TP}_{\rm max}$ (up to $3.8\times 10^8
$K) with respect to their solar-like metallicity counterpart. At
[Fe/H]=-2.15 the absolute maximum temperature is reached in the 5
M\odos model and not in the 6 M\odos one. In this case, a larger
mass of the H-exhausted core (which implies higher T$^{\rm
TP}_{\rm max}$) can not compensate the decrease of the duration of
the interpulse phases.

\section{ph-FRUITY: a new web physical interface}\label{phf}

The FRUITY database \citep{cri11} is organized under a relational
model through the MySQL Database Management System. Its Web
interface allows users to submit the query strings resulting from
filling out appropriately the fields to the managing system,
specifying the initial mass, metallicity and rotational velocity.
Up to date, FRUITY was including our predictions for the surface
composition of AGB stars and the stellar yields they produce. For
each model, different types of Tables can be downloaded (elemental
and isotopic surface compositions; net and total yields;
\s-process indexes). In this work, we add a new module
(ph-FRUITY), containing the physical quantities of interest
characterizing AGB models. The downloadable quantities (given for
each Thermal Pulse, with and without TDU) are: the absolute age,
the duration of the previous interpulse phase ($\Delta t_{\rm
ip}$), the total mass (M$_{\rm tot}$), the mass of the H-exhausted
core (M$_{\rm H}$), the dredged up mass ($\delta $M$_{\rm TDU}$),
the $\lambda$ quantity, the maximum temperature attained at the
bottom of the convective zone generated by the TP (T$^{\rm
TP}_{\rm max}$), the mean bolometric magnitude of the previous
interpulse period ($M_{\rm bol  }=4.75-2.5 * \log$ L/L$_\odot$),
the logarithm of the mean surface temperature of the previous
interpulse period (log T$_{\rm eff}$) and, the logarithm of the
mean surface gravity of the previous interpulse period (log
$g$)\footnote{Those quantities are weighted averages on time.}.
Note that we stop the calculations once the TDU has ceased to
operate. However, the core mass continues to grow up to the nearly
complete erosion of the convective envelope by the strong stellar
winds. Such an occurrence is accounted for by providing a set of
key extrapolated physical quantities (labelled as \emph{EXTRA}).
First, we calculate the mass lost in the wind and the growth of
the H-exhausted core during the previous interpulse phase. Then we
extrapolate them by means of a 5$^{th}$ order polynomial. Then, we
derive other tabulated quantities ($\Delta t_{\rm ip}$, $M_{\rm
bol}$, log T$_{\rm eff}$ and, log $g$). Following the original
FRUITY philosophy, those quantities can be downloaded in a
``Multiple case format'' (the query returns multiple tables,
depending on the number of selected models) or in a ``Single case
format'' (the query returns a single table containing all the
selected models).

\section{The TP-AGB phase (II): nucleosynthesis}\label{agbnuc}

The nucleosynthesis occurring during the TP-AGB phase is extremely
rich. In fact, many types of nuclear processes are at work,
including strong force reactions (proton captures, neutron
captures, $\alpha$ captures) and weak force reactions ($\beta$
decays, electron captures). Nearly all the isotopes in the
periodic table are affected, apart from Trans-uranic species.
\begin{figure*}[tpb]
\centering
\includegraphics[width=\textwidth]{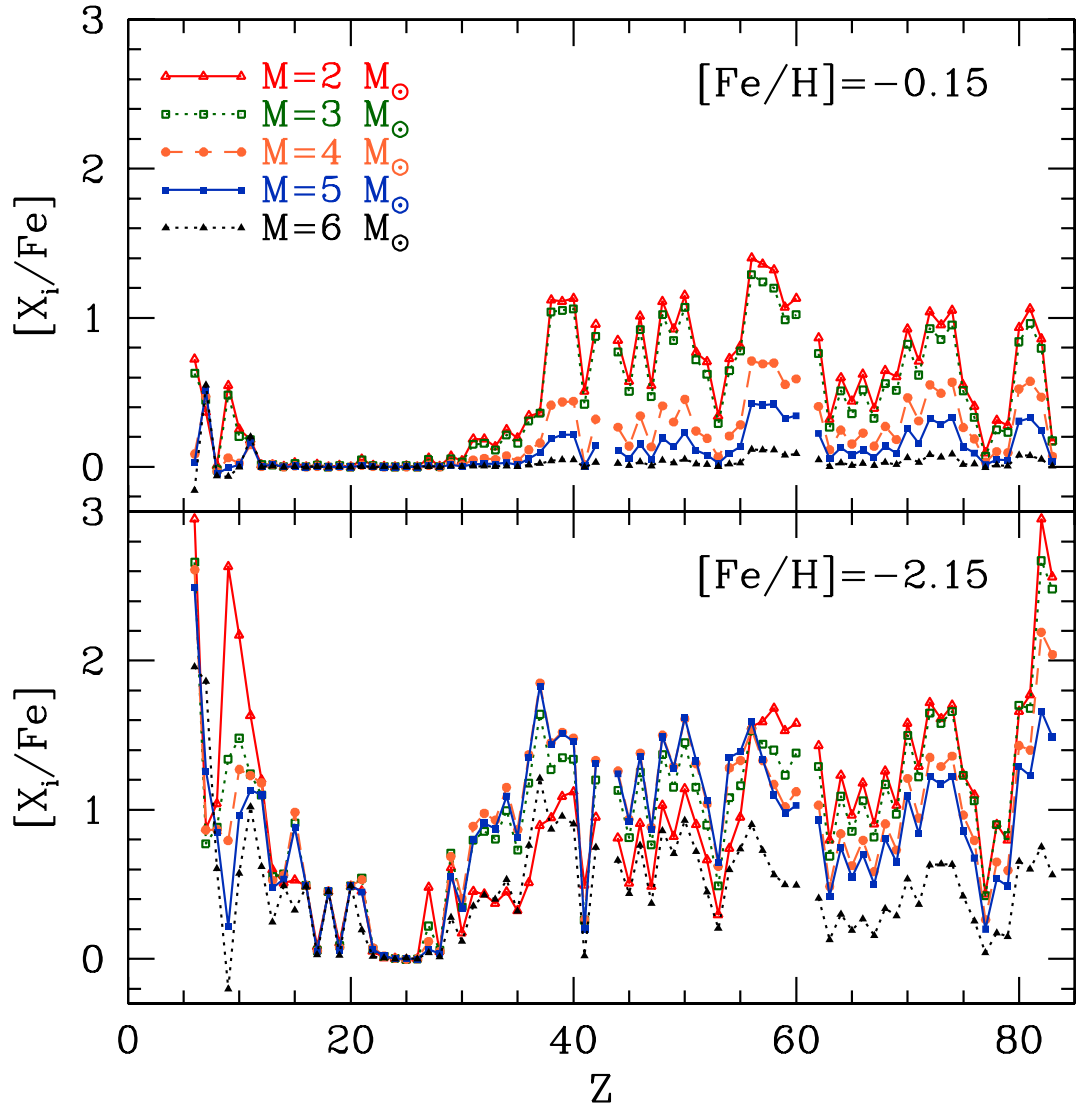}
\caption{Final element surface distribution for selected masses at
solar-like metallicity (upper panel) and at low metallicity (lower
panel). } \label{fig12}
\end{figure*}
\begin{figure*}[tpb]
\centering
\includegraphics[width=\textwidth]{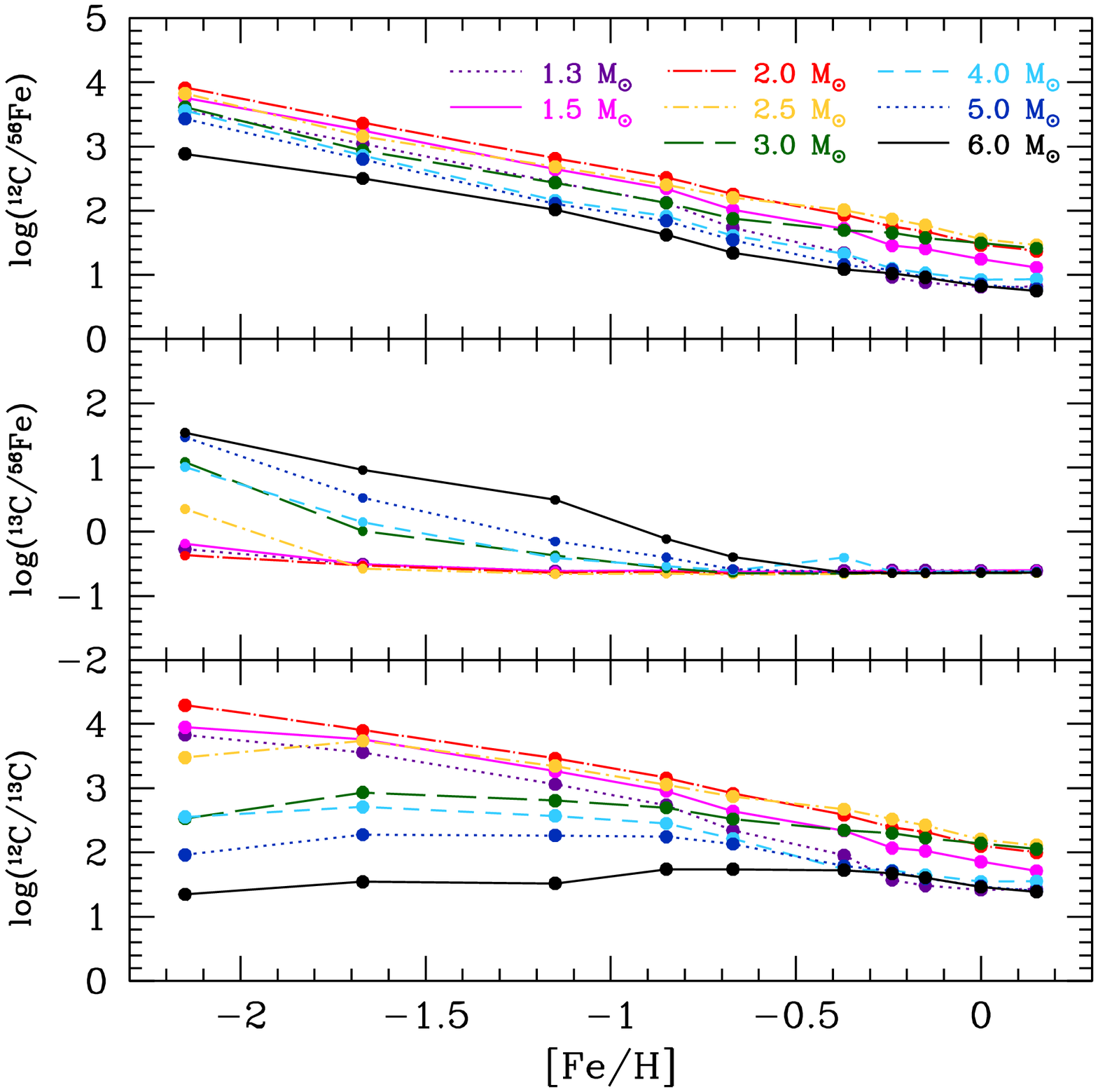}
\caption{Final surface distributions of $^{12}$C (upper panel) and
$^{13}$C (intermediate panel) with respect to $^{56}$Fe as a
function of the initial metallicity for the whole FRUITY set.
Final surface $^{12}$C/$^{13}$C ratios are also reported (lower
panel).} \label{fig13}
\end{figure*}
The nucleosynthetic details related to the evolution of low mass
TP-AGB stars have been already presented in \cite{cri09} and
\cite{cri11}. As we already stressed, the final surface abundances
and, consequently, the net yields are slightly smaller with
respect to data presented in those two papers, due to a previous
underestimation of the opacities in the most external layers of
the star. The proper opacities imply lower surface temperatures
and, thus, higher mass-loss rates. As a consequence, models
experience a reduced number of TDU episodes. However, \s-process
indexes are not affected by this problem (see \S \ref{agbnuc})
and, thus, most of the conclusions derived in \cite{cri11} are still valid.\\
In Figure \ref{fig12}, we report the final surface chemical
distributions\footnote{In the usual spectroscopic notation:
[X$_i$/Fe]= log
    (N(X$_i$)/N(Fe))$_{*}$ - log (N(X$_i$)/N(Fe))$_\odot$. Models with
initial masses M=1.3, M=1.5 and M=2.5 M\odos are omitted for
clarity.} for the whole mass range and two selected metallicities
(Z=$10^{-2}$: upper panel; Z=$2.4\times 10^{-4}$: lower panel). At
large $Z$, we notice a net production of carbon in LMS-AGBs only.
The 4 and 5 M\odos models present a slight final surface increase,
while the 6 M\odos model destroys it (due to the occurrence of the
FDU, the SDU and to a low TDU efficiency). At low metallicity all
models show a consistent production of $^{12}$C. In Figure
\ref{fig13}, we report the final surface number ratios
N($^{12}$C)/N($^{56}$Fe) and N($^{13}$C)/N($^{56}$Fe) as a
function of the initial metallicity for the whole FRUITY set
(upper panel and intermediate panel, respectively). There is a
clear increase in the $^{12}$C production as the initial iron
content decrease. The same is not true for $^{13}$C, which is
efficiently synthesized by the more massive AGBs and for
[Fe/H]$<$-0.5 only. The ($^{13}$C/$^{56}$Fe) ratio is in fact
nearly flat for models with M$<$2.5 M\odo. For larger masses (5-6
M\odo), instead, there is a net $^{13}$C production, due to the
simultaneous occurrence of HBB and H-TDU. The corresponding
increase of $^{14}$N can be visualized in Figure \ref{fig14}
(upper panel), as well as the decrease of the $^{12}$C/$^{13}$C
ratio (lower panel of Figure \ref{fig13}). We remind that the
$^{13}$C production is affected by the presence
of non-canonical mixing during the RGB phase, as already stressed in \S \ref{preagb}.\\
The solar oxygen abundance mainly consists of $^{16}$O. Its
production is null at large metallicities, while there is a net
production for the whole mass range at low $Z$, thanks to TDU
episodes, which mix to the surface the $^{16}$O. The latter is
synthesized by the $^{12}$C($\alpha$,$\gamma$)$^{16}$O reactions
during TPs and, lo a lesser extent, by the \ctan reaction during
the radiative \ct burning. The other oxygen isotopes ($^{17}$O and
$^{18}$O) exhibit completely different behaviors. The $^{17}$O
abundance results from the equilibrium between the production
channel ($^{16}$O(p,$\gamma$)$^{17}$F($\beta^+$)$^{17}$O nuclear
chain) and the destruction one ($^{17}$O(p,$\alpha$)$^{14}$N
reaction). Its surface abundance depends on the depth of
convection in regions with an $^{17}$O profile, i.e. those
experiencing an incomplete CNO burning. For a fixed [Fe/H], we
find the highest $^{16}$O/$^{17}$O for the lowest masses (middle
panel in Figure \ref{fig14}), thus confirming the values already
reported in the literature (see e.g. \citealt{lebze15}). The
complex interplay between mixing and burning does not allow to
identify a common behavior with the metallicity. However, the
models show slightly larger $^{16}$O/$^{17}$O ratios at large
metallicities. The more neutron rich oxygen isotope ($^{18}$O)
behaves very differently (see lower panel of \ref{fig14}). This
isotope is mainly produced by the
$^{14}$N($\alpha$,$\gamma$)$^{18}$F($\beta^+$)$^{18}$O nuclear
chain, while it is destroyed by the $^{18}$O(p,$\alpha$)$^{15}$N
and $^{18}$O($\alpha$,$\gamma$)$^{22}$Ne reactions. The
$^{16}$O/$^{18}$O ratio is nearly constant for all masses and
[Fe/H]$\ge$-1.15. At low metallicities, its increase is due to the
dredge up of primary $^{16}$O, as explained before. Thus, in our
models $^{18}$O is basically untouched. This would not be the
case, if we would take into consideration the effects induced by
non convective mixing during the RGB and the AGB phases. In fact,
it has been demonstrated that the inclusion of this kind of mixing
strongly affects the surface $^{18}$O abundance of low mass stars
and that this is needed to fit laboratory measurements of oxygen
isotopic ratios in pre-solar SiC grains (see
\citealt{palmerini2011}). Those small dust particles are trapped
in primitive meteorites and currently provide the most severe
constraints to AGB nucleosynthesis (see e.g.
\citealt{liu14,liu15}).
\\
The fluorine nucleosynthesis is extremely complex, since it
involves both neutron and proton captures
\citep[see][]{abia09,abia10,abia11}. $^{19}$F is very sensitive to
a variation of the initial stellar mass (see lower panel of Figure
\ref{fig12}). Its production basically depends on the amount of
$^{15}$N in the He-intershell, which in turn is correlated to the
amount of $^{13}$C in the ashes of the H-burning shell, as well as
in the $^{13}$C pocket \citep[see the discussion in ][]{cri14}. In
IMS-AGBs, fluorine production is strongly suppressed due to the
reduced contribution from the radiative \ct burning and from the
increased efficiency of $^{19}$F destruction channels (the
$^{19}$F(p,$\alpha$)$^{16}$O reaction and,
above all, the $^{19}$F($\alpha$,p)$^{22}$Ne reaction).\\
Neon is enhanced in all the models experiencing TDU, due to the
dredge up of the freshly synthesized $^{22}$Ne during TPs via a
double $\alpha$ capture on the abundant $^{14}$N. Its abundance
directly affects the $^{23}$Na nucleosynthesis.  In LMS-AGBs,
sodium can be synthesized through proton captures during the
formation of the \ct pocket, as well as through neutron captures
during both the radiative burning of the \ct pocket and the
convective $^{22}$Ne-burning in the convective shells generated by
TPs (see \citealt{cri09}). This leads to a notable $^{23}$Na
surface enhancement, in particular at low metallicities. In more
massive stars, the sodium nucleosynthesis is affected by HBB
\citep[see, e.g.][]{veda06,kakka14}. In our models, we find a
slight increase of the $^{23}$Na surface abundance directly
correlated to HBB, which is mildly
activated in the more massive low Z models.\\
\begin{figure*}[tpb]
\centering
\includegraphics[width=\textwidth]{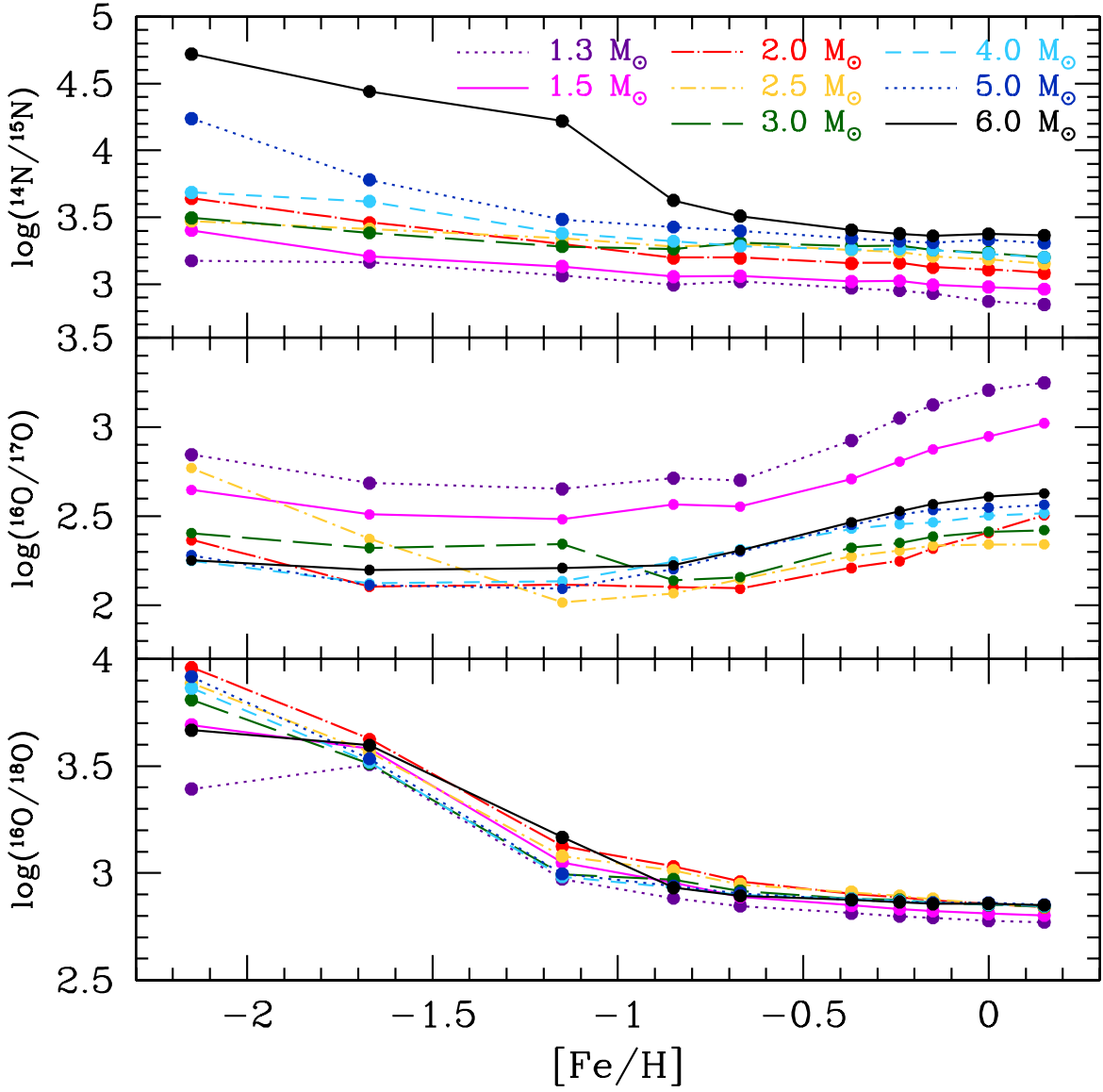}
\caption{Final surface distributions of the $^{14}$N/$^{15}$N
ratio (upper panel), the $^{16}$O/$^{17}$O ratio (intermediate
panel) and, the $^{16}$O/$^{18}$O ratio (lower panel).}
\label{fig14}
\end{figure*}
\begin{figure*}[tpb]
\centering
\includegraphics[width=\textwidth]{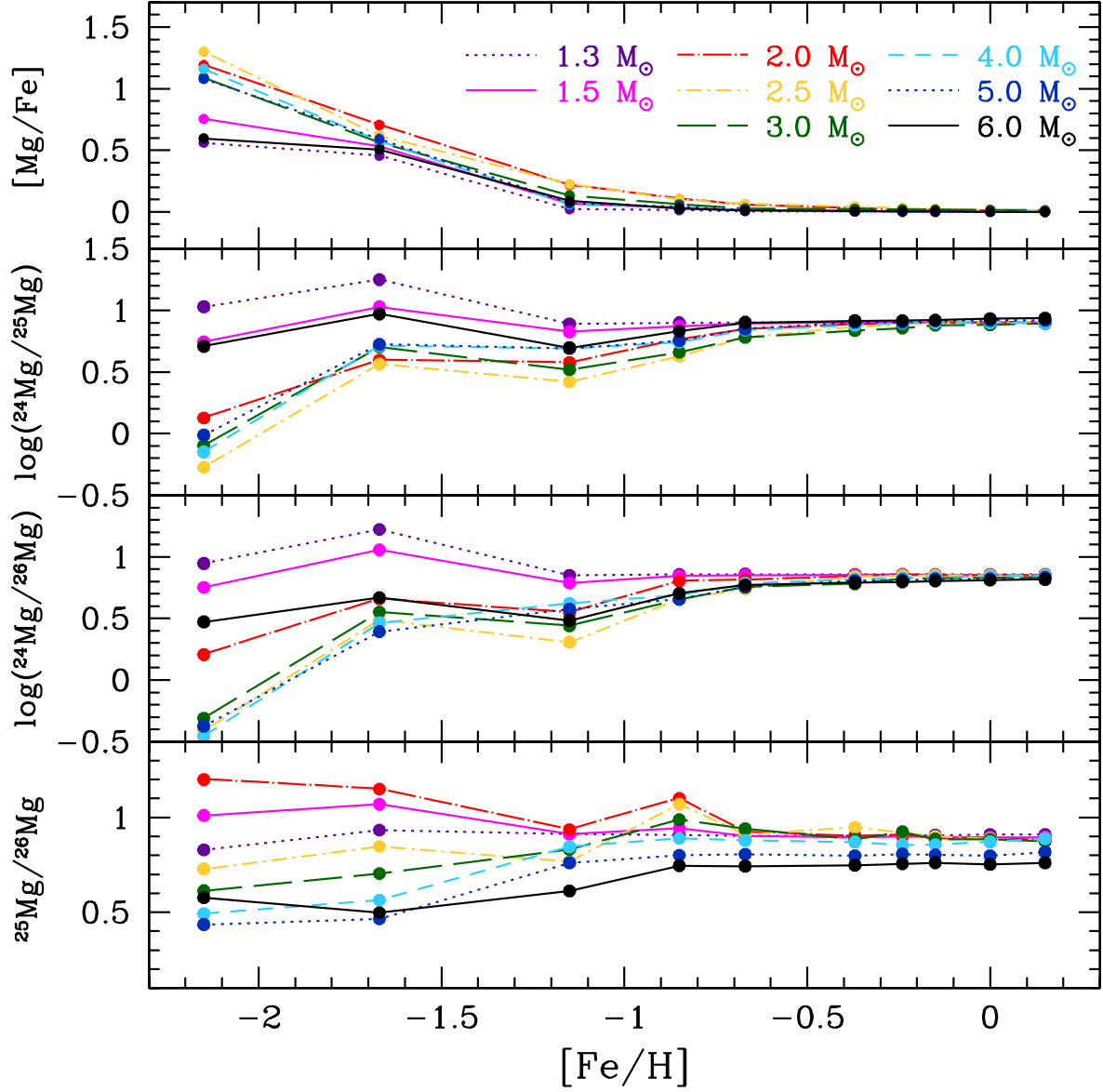}
\caption{Final surface [Mg/Fe] (upper panel), $^{24}$Mg/$^{25}$Mg
and $^{24}$Mg/$^{26}$Mg ratios (intermediate panels) and,
$^{25}$Mg/$^{26}$Mg ratio (lower panel).} \label{fig15}
\end{figure*}
\begin{figure*}[tpb]
\centering
\includegraphics[width=\textwidth]{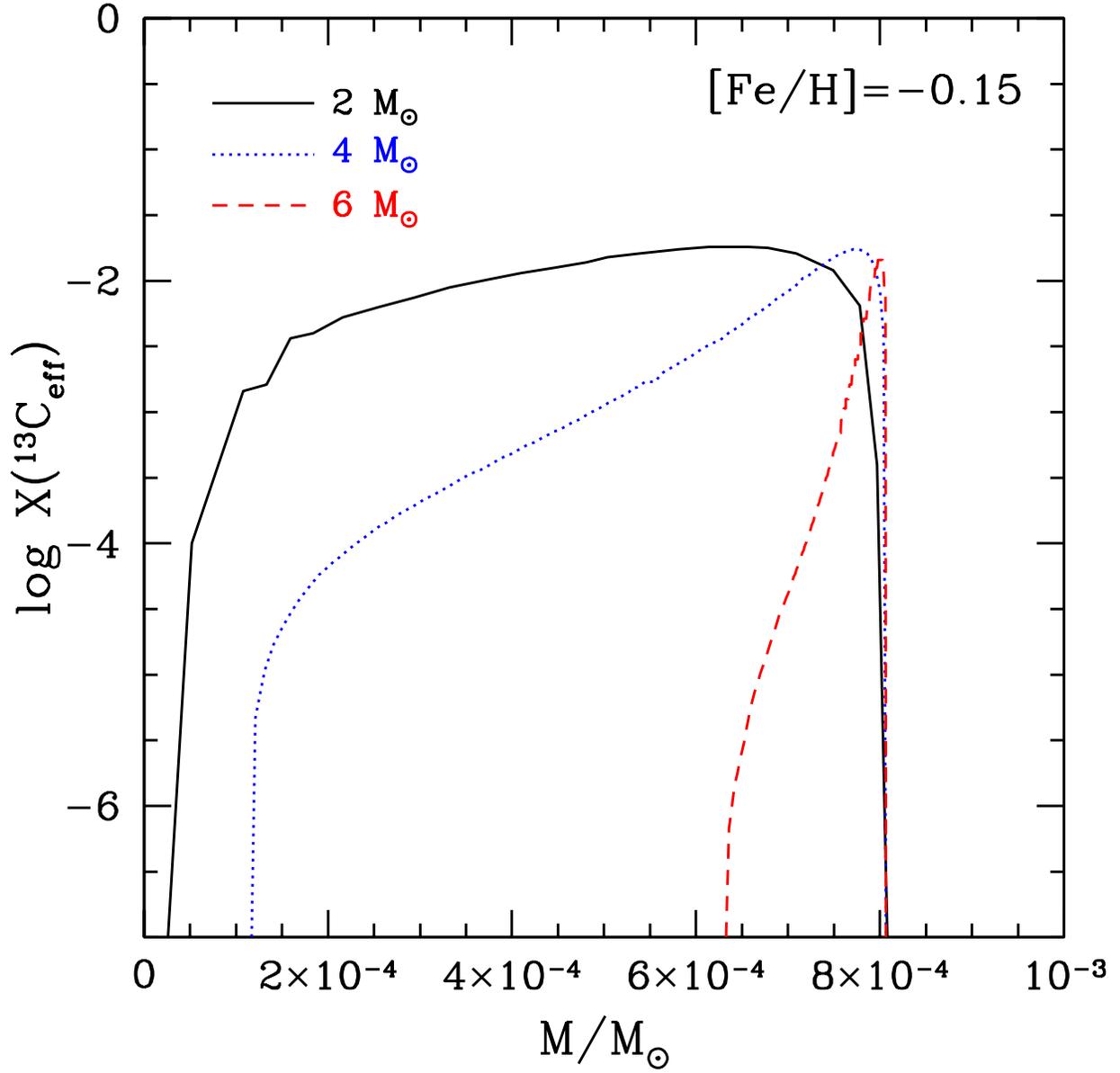}
\caption{Mass extension of the {\it effective} \ct in the pockets
after the second TDU of the 2.0, 4.0 and, 6.0 M\odos models with
$Z=10^{-2}$. See text for details.} \label{fig16}
\end{figure*}
In our models, magnesium is enhanced at low metallicities, due to
an increased production of $^{25}$Mg and $^{26}$Mg, via the \nean
and the \neag reactions (see upper panel of Figure \ref{fig15}).
We find a considerable Mg overabundance at [Fe/H]=-2.15 only (we
remind that for [Fe/H]$\le$-1.67 we adopt an $\alpha$-enhanced
initial mixture). At low metallicities both the
$^{24}$Mg/$^{25}$Mg and the $^{24}$Mg/$^{26}$Mg are lower than
solar (intermediate panels of Figure \ref{fig15}). Both quantities
show a minimum for models with initial mass M$\sim$3 M\odo.
Exceptions are represented by the less massive models (1.3 M\odos
and 1.5 M\odo), in which the marginal activation of both the \nean
and the \neag reactions does not compensate the initial $^{24}$Mg
enhancement. At intermediate-to-high metallicities the final
surface $^{26}$Mg/$^{25}$Mg ratio of our models is nearly constant
(and close to the solar value). At low [Fe/H], instead, it depends
on the initial mass, the 2 M\odos showing the maximum value and
the 5 M\odos the minimum one (see lower panel of Figure
\ref{fig15}). During TPs, in the massive models the neutron
density is large and $^{25}$Mg behaves as a neutron poison, thus
feeding $^{26}$Mg.

The contribution to the overall nucleosynthesis from the \ctan
reaction is strongly correlated to the initial mass of the model.
In \cite{cri09}, we showed that the mass extension of the \ct
pocket decreases with the TDU number (see also Figure 1 in
\citealt{cri11}), implying a progressive reduction of the
\s-process efficiency as the star evolves along the AGB. The
larger the core mass, the lower the \ct size. Thus, \ct pockets in
IMS-AGB models are definitely thinner with respect to those found
in LMS-AGBs, due to their definitely larger core masses. In Figure
\ref{fig16}, we compare the profiles of the {\it effective} \ct
(defined as X($^{13}$C$_{\rm
eff}$)~=~X($^{13}$C)~-~13/14~*~X($^{14}$N)) in the pocket after
the $2^{nd}$ TDU for three different models (2.0, 4.0 and, 6.0
M\odo) with $Z=10^{-2}$. This quantity takes into account the
poisoning effect of $^{14}$N (via the $^{14}$N(n,p)$^{14}$C
reaction) and, thus, provides a better estimate of the neutrons
effectively contributing to the synthesis of heavy elements (see
\citealt{cri09}).
\begin{figure*}[tpb]
\centering
\includegraphics[width=\textwidth]{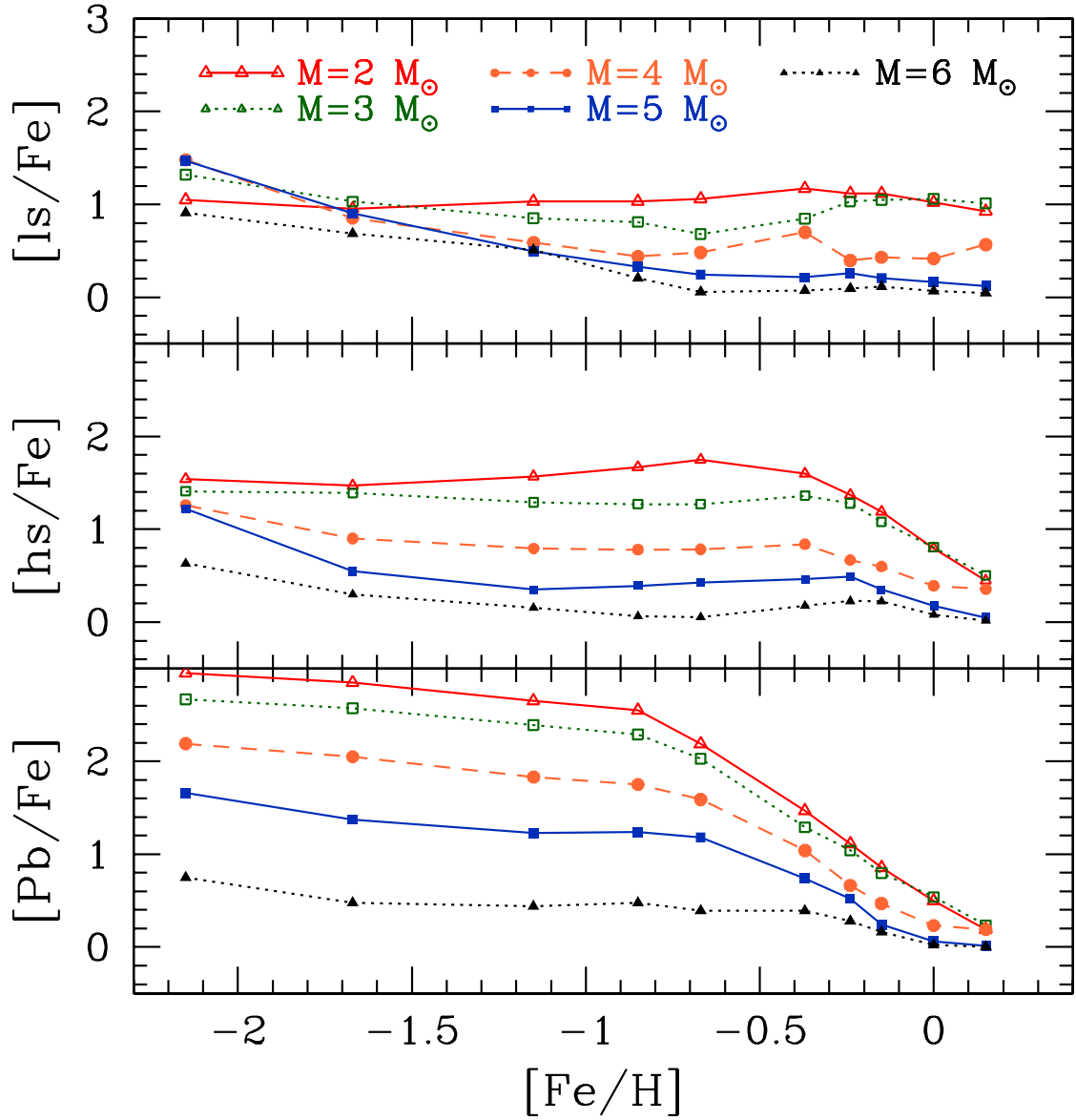}
\caption{Final surface \s-process enhancements: ls component
(upper panel), hs component (middle panel) and, lead (lower
panel).} \label{fig17}
\end{figure*}
\begin{figure*}[tpb]
\centering
\includegraphics[width=\textwidth]{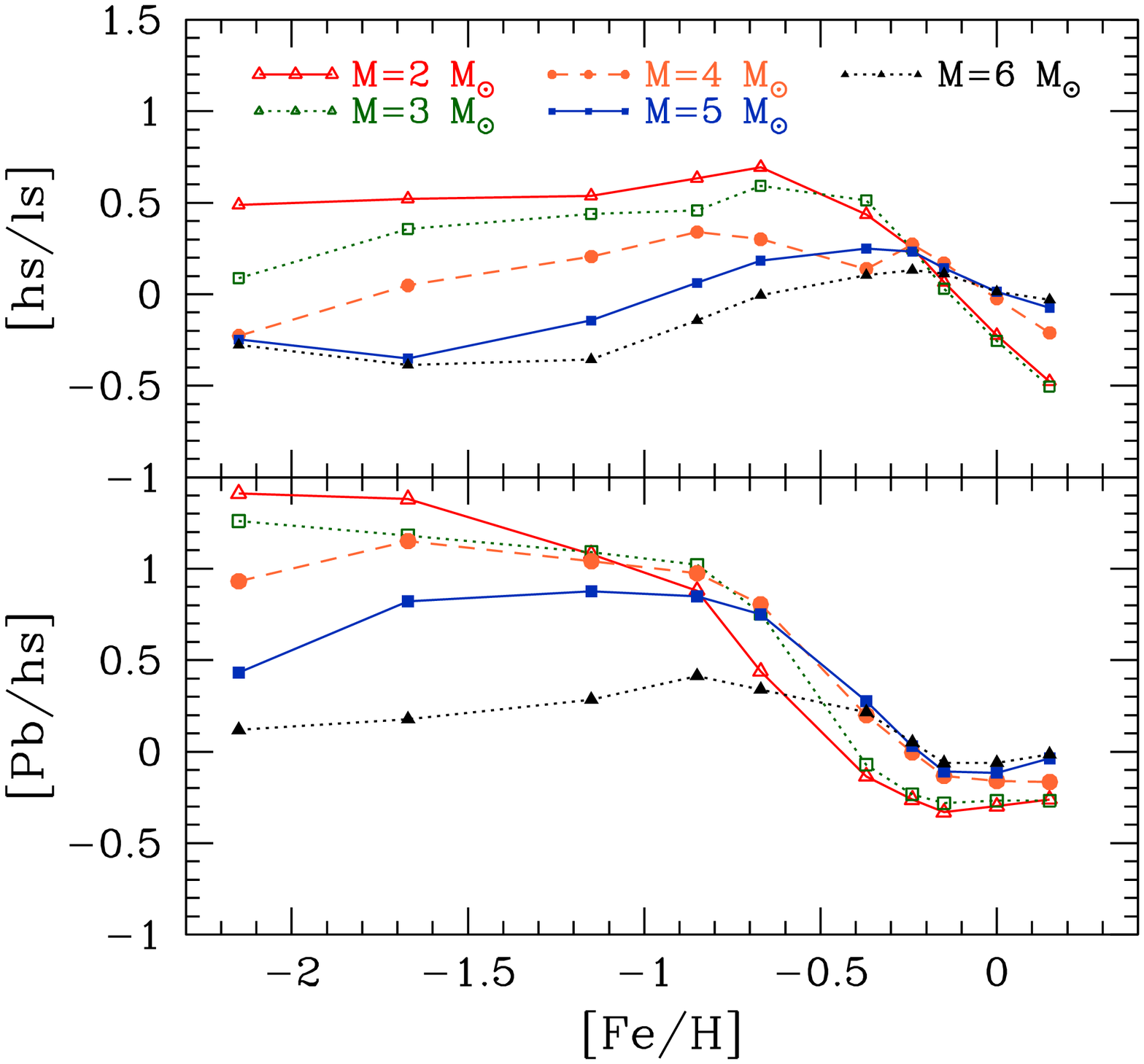}
\caption{Final surface \s-process indexes: [hs/ls] (upper panel)
and [Pb/hs] (lower panel). } \label{fig18}
\end{figure*}
In Figure \ref{fig16}, the pockets have been manually shifted in
mass, while the zero point of the abscissa is arbitrary. The
pocket found in the 6 M\odos model is four times smaller than that
found in the 2 M\odos model, due to the shrinking of the
He-intershell with increasing initial stellar mass. Moreover, the
integrated amount of effective \ct in the pocket decreases by more
than a factor of 3, passing from the 2 M \odos model ($\sum
^{13}$C$_{\rm eff}=7.8 \times 10^{-6}$ M\odo) to the 4 M\odos
model ($\sum ^{13}$C$_{\rm eff}=2.2 \times 10^{-6}$ M\odo) and is
reduced by another factor of 7 in the 6 M\odos model ($\sum
^{13}$C$_{\rm eff}=3.1 \times 10^{-7}$ M\odo). This fact has
obvious consequences on the production of \s-process elements. The
main neutron source in LMS-AGBs is the \ctan reaction
\citep{ga98,stra06}. In those stars, a marginal contribution comes
from the partial activation of the \nean reaction. In
\cite{cri11}, we extensively described the final \s-process
surface distributions for AGB stars in this mass range. Their
heavy element distributions show a progressive drift to heavier
nuclei as the metallicity decreases. This is due to the fact that
the neutron source (the \ctan reaction) is of primary origin,
while the seeds ($^{56}$Fe) scale with metallicity. As a
consequence, the lower the initial iron content, the larger the
neutron-to-seed ratio. In IMS-AGBs stars, this scheme still holds,
but with important differences. As already stressed in
\cite{stra14}, these objects develop larger temperatures at the
base of convective shells during TPs, thus efficiently activating
the \nean reaction. Moreover, the contribution from the \ctan
reaction is lower, due to thinner \ct pockets, as shown before. A
clear sign of the \nean activation is reflected in the rubidium
surface enhancement (see lower panel of Figure \ref{fig12}). It
comes from the large production of $^{87}$Rb, which is by-passed
during the radiative \ct burning, due to the branchings at
$^{85}$Kr and, to a lesser extent, at $^{86}$Rb \citep{stra14}.
Since the neutron exposure during the \nean episode is lower with
respect to that of the \ctan reaction, we also expect an overall
reduction of \s-process overabundances, in particular for the
second and third peak of the \s-process. This is confirmed by
Figure \ref{fig17}, in which we report the behavior of the three
\s-process peaks as a function of the metallicity. The
corresponding data are tabulated in Tables \ref{tab_ls},
\ref{tab_hs} and, \ref{tab_pb}. IMS-AGBs show definitely lower
surface enhancements for the hs component (intermediate panel) and
lead (lower panel). Note that this is also due to the reduced TDU
efficiency characterizing those models (see \S \ref{agb}).
However, at low metallicities the ls component (upper panel) of
these models is comparable to that of less massive objects, thus
demonstrating that the \nean reaction is efficiently at work. In a
convective environment, neutrons cannot be released and piled up
to synthesize the heaviest elements. This is particularly evident
for lead, whose production is hampered in IMS-AGBs. The
corresponding \s-process indexes [hs/ls] and [Pb/hs] are reported
in Figure \ref{fig18}. The corresponding data are tabulated in
Tables \ref{tab_hsls} and \ref{tab_pbhs}. As already anticipated
in previous Sections, we find a general reduction of the surface
enhancements with respect to yields from \cite{cri11}. However,
the [hs/ls] and [Pb/hs] indexes remain almost unaltered because in
LMS-AGBs models those quantities are nearly independent on the
evolutionary stage along the AGB, provided that the \s-process
enhancement is sufficiently large. As already stressed, this
derives from the fact that the first \ct pockets (the largest
ones) are those governing the whole nucleosynthesis. \cite{cri15}
recently computed the galactic chemical evolution of \s-only
isotopes and found that FRUITY models predict too large abundances
for those nuclei. Thus, the reduction of LMS-AGBs yields we
discussed in this paper may lead to a better agreement with the
observed solar \s-only distribution. This problem will be
addressed
in a forthcoming paper. \\
The behavior of the [hs/ls] index strongly depends on the initial
stellar mass. In fact, the efficient activation of the \nean
reaction and the reduced contribution from the \ctan reaction
lead, in IMS-AGBs, to low [hs/ls] and [Pb/hs] indexes (see Figure
\ref{fig18}). Another striking difference with respect to LMS-AGBs
is that our massive AGBs do not reach an asymptotic value in the
surface ratio between \s-process peaks. This can be appreciated in
Figure \ref{fig19}, where we plot the [hs/ls] (upper panel) and
[Pb/hs] (lower panel) as a function of the TDU number for selected
masses with $Z=2.3\times 10^{-4}$. In LMS-AGBs, those quantities
rapidly grow up to an asymptotic value and then remain basically
frozen (see the 2 M\odos curve). For larger masses, instead, both
indexes reach a maximum and then start decreasing. This behavior
is determined by the additional contribution from the \nean
reaction, which is able to synthesize ls elements, but not to
produce the hs ones. Unlike the \ct pockets, the {\it imprint} of
this neutron source progressively emerges with the TDU number.
This is due to the fact that the largest temperatures are attained
at the base of the convective shells generated by TPs toward the
end of the AGB phase (see Figures \ref{fig10} and \ref{fig11}).
The larger the initial mass, the larger the TDU number needed to
achieve such an asymptotic regime\footnote{Note that the 6 M\odos
model experiences more than 80 TDUs, but its [hs/ls] and [Pb/hs]
are practically constant after the 50th TDU.}. The [Pb/hs] does
not depend on the activation of the \nean source, but only on the
\ctan one. Thus, the difference between LMS and IMS is less
evident. Obviously, the larger the initial mass, the lower the
[Pb/hs] value, due to the decreasing contribution from the \ct
pockets.
\begin{figure*}[tpb]
\centering
\includegraphics[width=\textwidth]{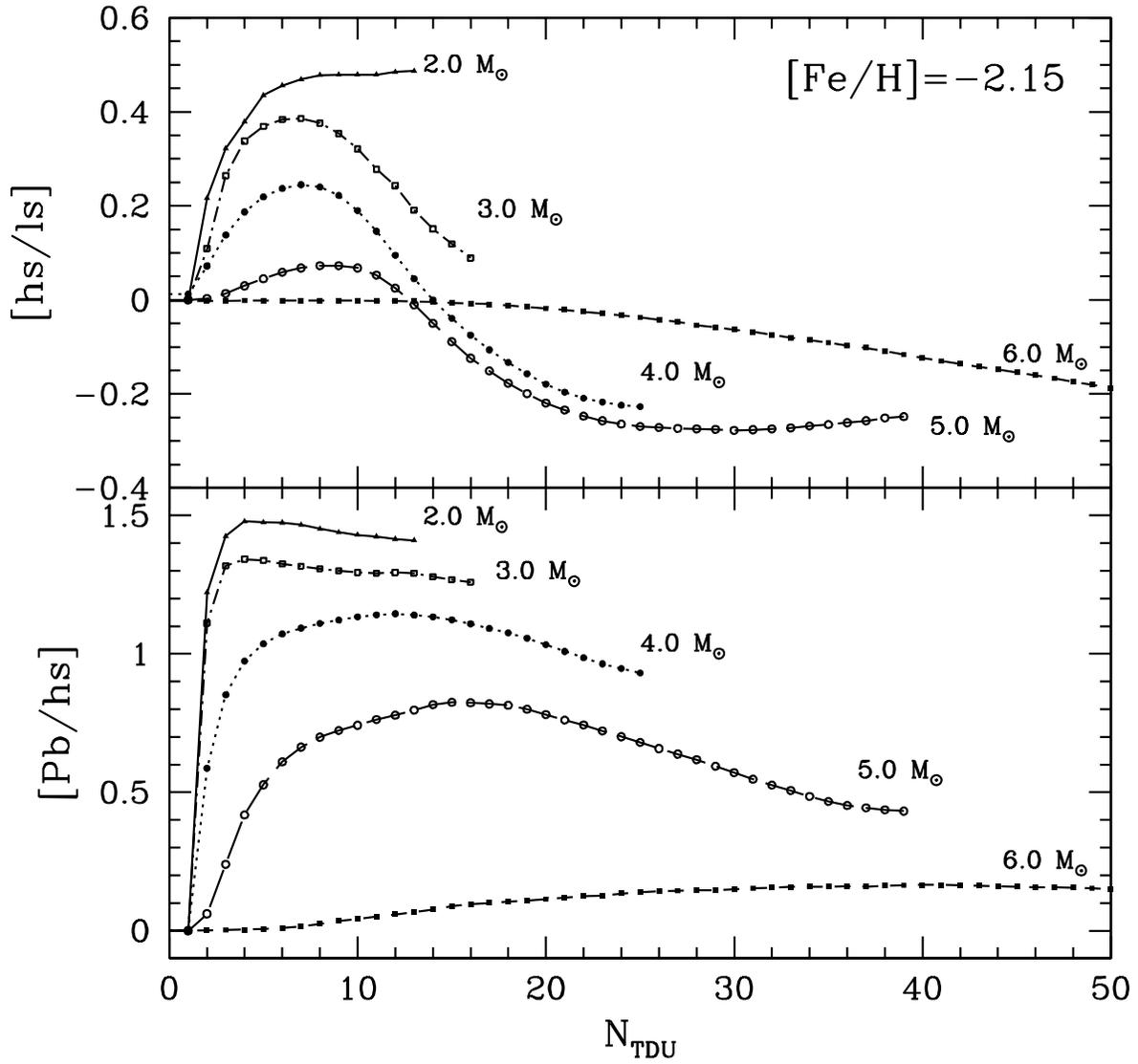}
\caption{[hs/ls] (upper panel) and [Pb/hs] (lower panel) as a
function of the TDU numbers for selected models with
[Fe/H]=-2.15.} \label{fig19}
\end{figure*}

\section{Discussion and Conclusions} \label{conclu}

\begin{figure*}[tpb]
\centering
\includegraphics[width=\textwidth]{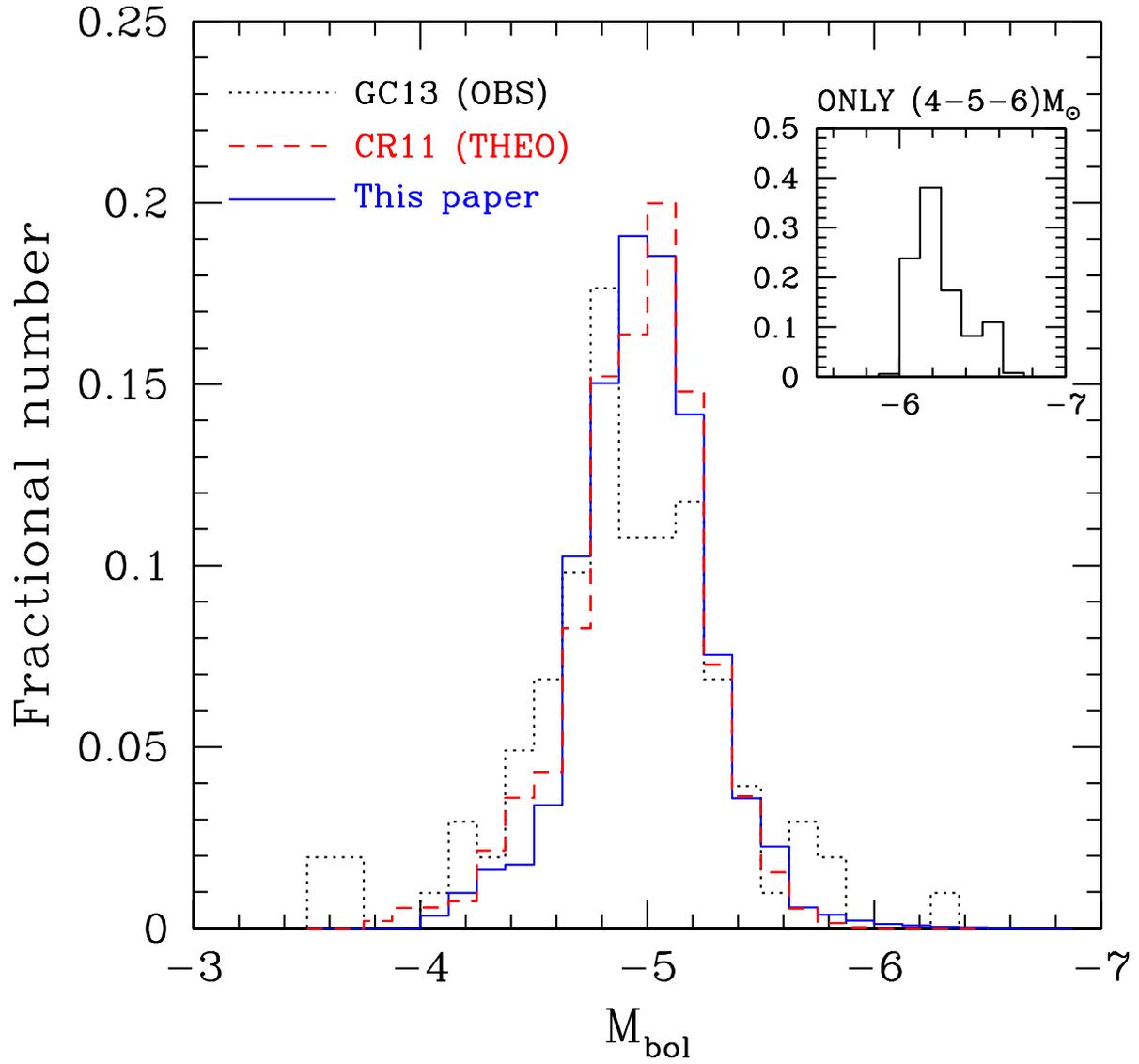}
\caption{Our new theoretical Luminosity Function of Carbon Stars
(solid curve) compared to our previous estimate (dashed curve;
CR11: \citealt{cri11}) and to observations (dashed curve; GC13:
\citealt{guanda13}).} \label{fig20}
\end{figure*}

\begin{figure*}[tpb]
\centering
\includegraphics[width=\textwidth]{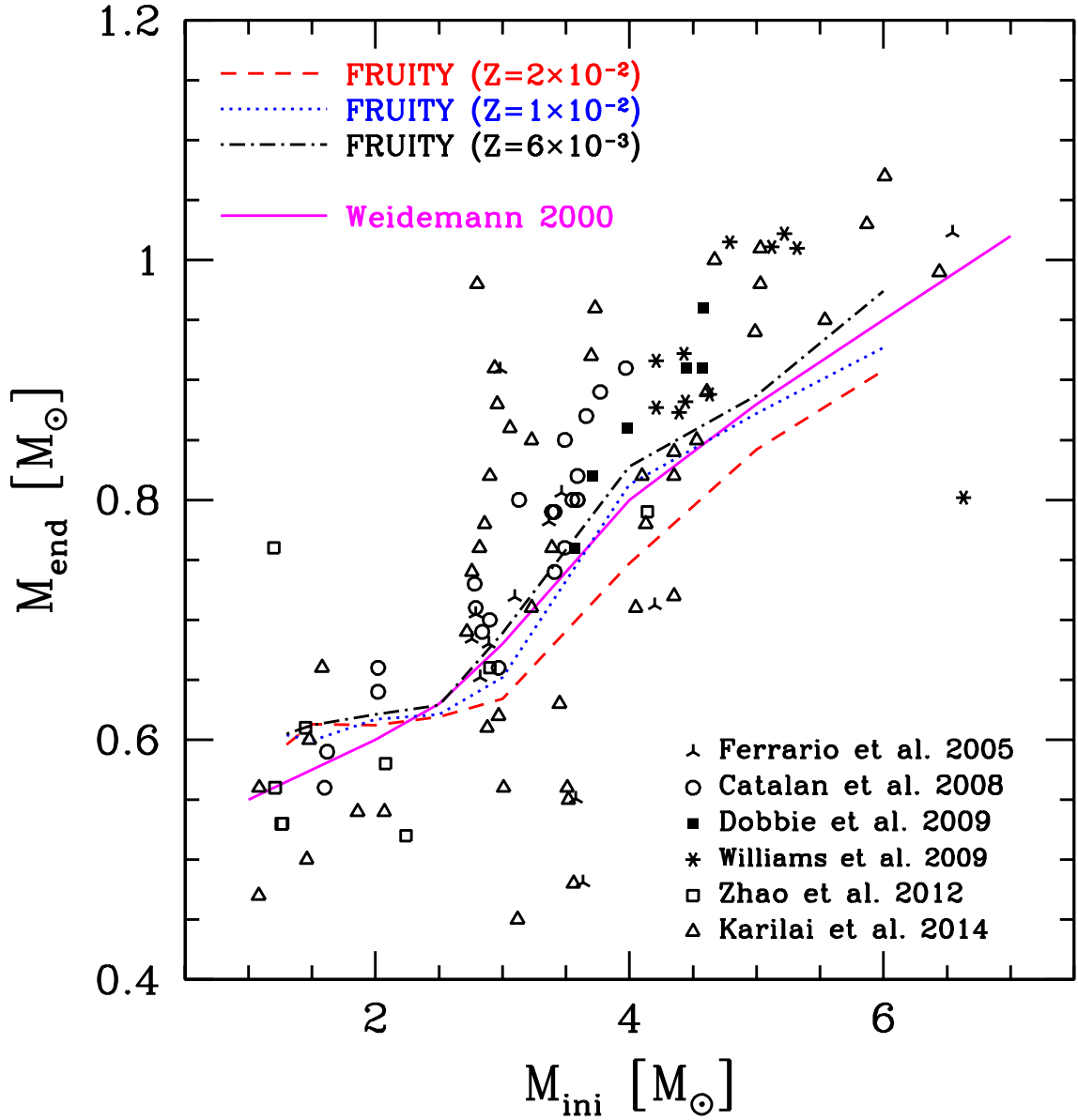}
\caption{FRUITY models Initial-to-final mass relations for
selected metallicities, compared to the semi-empirical relation
from \cite{weidemann} and to Open Clusters observations.}
\label{fig21}
\end{figure*}

A detailed comparison between our LMS-AGB models and those from
other groups have been already presented in \cite{cri11} (see
Section 5 of that paper). A similar analysis, but for IMS-AGBs,
can be found in \cite{ventura13}, who compared their models with
those published by \cite{kakka10}, as well as in \cite{fi14}, who
made a comparison with a subset of models presented by
\cite{stra14}. Notwithstanding, in Table \ref{tab_compa} we report
key quantities related to our 5 M\odos model with $Z=1\times
10^{-3}$ compared to similar available models in the literature
\citep{fi14,veda08}. Note that there are other published papers on
IMS-AGBs; however, they concentrate on different mass regimes
(e.g. \citealt{siess07,dohe15}) or present sets for a single
metallicity (\citealt{he04}). In general, this kind of comparison
is not straightforward, since evolutionary codes significantly
differ in the adopted input physics (the treatment of convection
and convective borders; the mass-loss rate; the initial chemical
distribution; the opacities; the equation of state; the nuclear
network; etc). For instance, a direct comparison between our
models and those presented by \cite{veda08} is difficult because
not only the treatment of convective borders is different, but
also the theoretical recipe to model convection is not the same
(we use the MLT formulation by \citealt{cox68}, while
\citealt{veda08} adopt the Full Spectrum of Turbulence of
\citealt{cama91}). Other differences are the adopted mass-loss
rate (\cite{veda08} use a calibrated version of the
\cite{bloecker} mass-loss formula) as well as the initial chemical
distribution (\cite{veda08} adopt the solar mixture by \cite{gs98}
with an $\alpha$ element enhancement of 0.4). From an inspection
of Table \ref{tab_compa} it turns out that our model shows a HBB
shallower than the model by \cite{veda08}. It has to be remarked
that in the models by \cite{veda08} mixing and burning are
coupled. As already stressed, we aim to verify the effects of such
a coupling in our models in a future work. Actually, due to the
adopted input physics (MLT for convection and mass-loss rate
calibrated on galactic AGB stars), our model should be more
similar to that of \cite{fi14}. However, also in this case the
differences are notable. At odds with our model, the 5 M\odos
model of \cite{fi14} experience a stronger HBB (as testified by
the large temperatures attained at the base of the convective
envelope during interpulse periods). In our models, we test the
effects of changing the efficiency of mixing (we vary the free
parameter of the MLT or the $\beta$ parameter governing the
convective velocity profile at the base of the convective
envelope), the mixing scheme (by assuming instantaneous mixing in
the envelope), the treatment of opacities at the border of the
convective envelope, the adopted mass-loss rate (we run a model
without mass-loss and let it to evolve to larger core masses) or
the equation of state (EOS; we substitute our
treatment\footnote{\cite{pd02} for T$> 10^6$ K and the Saha
equation for lower temperatures (see \citealt{stra88}).} by
adopting the OPAL EOS 2005 at high temperatures \citep  {opal} and
checking different transition temperatures to the low temperature
regime). Those test lead to variations of the $^{12}$C/$^{13}$C
ratio, but none of them shows a significant activation of HBB.
Thus, we are not able to explain such a discrepancy in the thermal
stratification of our models with respect to \cite{fi14}. Perhaps
the origin has to be searched in the dated EOS used by
\cite{fi14}. As reported in \cite{dohe15}\footnote{We suppose
\cite{fi14} use the stellar code matrix.}, the perfect gas
equation is adopted for fully ionized regions, the Saha EOS in
partially ionized regions (following the method of
\citealt{ba65}), while EOS from \cite{bt71} is used for
relativistic or electron-degenerate gas. However, a discussion on
the proper EOS to be used in AGB stellar models, as well on the
effects induced by adopting different EOS, is beyond
the goals of this paper.\\
More meaningful conclusions can be derived by comparing
theoretical models to observed quantities. As discussed in \S
\ref{agb}, the majority of our models present final C/O ratios
larger than 1. Their observational counterparts are C-rich stars,
whose Luminosity Function, which links a physical quantity (the
luminosity) with the chemistry (its surface carbon abundance),
represents a good test indicator for theoretical prescriptions. A
revision of the observational galactic Luminosity Function of
Carbon Stars (LFCS) has been recently presented by
\cite{guanda13}. Such a LFCS is plotted in Figure \ref{fig20}
(dotted histogram), together with the theoretical LFCS obtained
with models by \cite{cri11} (dashed histogram) and with models
presented in this paper (continuous histogram). With respect to
our previous estimate, we note a marginal shift to low
luminosities, as a consequence of the reduced TP-AGB lifetimes
caused by the erroneous treatment in the opacities of the most
external layers of the star. In the upper right corner of Figure
\ref{fig20} we report the LFCS derived by considering the
contribution of IMS-AGBs only. Those objects populate the high
luminosity tail of the LFCS. However, their contribution to the
whole distribution is practically negligible. Thus, at variance
with LMS-AGBs, the LFCS cannot be fruitfully used to constrain the
physical
evolution of IMS-AGBs.\\
Another interesting quantity, which can be used to constrain
theoretical models, is the initial-to-final mass relation. This
relation depends on the core mass of the H-exhausted core attained
at the end of the TP-AGB phase. As already recalled in \S
\ref{agb}, for the more massive models presented here the presence
of the SDU induces important variations in the mass of the
H-exhausted core. In Figure \ref{fig21} we compare a selection of
our models to the semi-empirical initial-to-final mass relation of
\cite{weidemann} as well as to observational data of Open Clusters
\citep{fe05,cat08,do09,wi09,zhao12,kali14}. In Table
\ref{tab_mcore} we report the final core masses of the whole
FRUITY set.
Our theoretical models agree well with the semi-empirical
initial-to-final mass relation of \cite{weidemann}, showing
however larger core masses for low mass objects. When looking to
observations, the situation becomes more complex. For a fixed
initial mass, observations present a rather wide spread up to
$\sim 0.5$ M\odo. Thus, firm conclusions cannot be drawn.
Moreover, it has to be taken into account that many observations
are indirectly affected by the uncertainty characterizing stellar
models. In fact, while the WD mass can be determined basing on
spectroscopic data, the initial mass is generally derived by
estimating the cluster age. This evaluation is done by means of
theoretical relations among mass, age and turn off luminosity.
Thus, the result depends on the physical recipe adopted to compute
the cluster isocrone. In Figure \ref{fig21} we report core masses
at the last computed model (i.e. when TDU ceases to operate). The
computing of the following evolutionary phases is made difficult
by the treatment of the most external layers. The final surface
chemistry remains frozen up to the WD phase, unless a very Late
Thermal Pulse occurs \citep{iben83,he11}. Note that only under the
hypothesis of a strong final super-wind episode, able to
instantaneously remove the whole remaining envelope, core masses
at the last TDU would coincide with WD masses. Alternatively, the
star experiences additional TPs without TDU up to the almost
complete erosion of the convective envelope. Then, we also
performed an extrapolation of the core mass. For models with an
initial mass larger than 3 M\odo, the differences in the core mass
between the last computed model and the extrapolated number are
neglibile (lower than 0.01 M\odo), while in the LMS regime they
become appreciable (up to 0.035 M\odo). The
extrapolated final masses are reported in the ph-FRUITY database.\\
Spectroscopic observations, able to constrain the evolution of
IMS-AGBs, are rare. The lack of C-stars in Magellanic Clouds (MCs)
with luminosities larger than M$_{\rm bol}$=-6 led \cite{wood83}
to the conclusion that CNO cycling is at work in upper AGB stars,
preventing them to become C-rich. Actually, it cannot be excluded
that C-rich stars with high luminosities exist, since they could
be embedded in an opaque dust rich cloud masking them to
observations. For instance, \cite{vanloon99} found giant C-rich
stars up to M$_{\rm bol}$=-6.7, thus demonstrating that in those
objects TDU is at work and that HBB is not efficient enough to
make those stars O-rich. More recently, a restricted sample of
galactic O-rich giant stars has been presented by \cite{gh06}.
Unfortunately, in the spectral regions under analysis, the adopted
synthetic spectra do not have the necessary resolution to
precisely fix the carbon, nitrogen, and oxygen abundances (as well
as the $^{12}$C/$^{13}$C ratio, see \citealt{gh07}). Moreover, for
galactic stars the determination of the absolute magnitude is
highly uncertain due to the difficulties in determining the
distance of those objects. Thus, apart from the derived C/O ratio
(less than 0.75) and the lithium abundance, they cannot be used to
constrain HBB. However, our models can be tested by checking their
\s-process elements abundances. Stars observed by \cite{gh06} have
been found extremely rubidium-rich and zirconium poor. This is at
odds with theoretical \s-process expectations. However, more
recently the same authors \citep{zamo14} re-analyzed 4 stars
demonstrating that the inclusion of a circum-stellar component
leads to definitely lower Rb surface abundances, without
appreciably modifying Zr data. Looking to their Table 1, we notice
that for the galactic sample all stars are compatible, within
errors, with null \s-process enrichments (for both Rb and Zr) but
IRAS 18429-1721, showing an appreciable Rb enrichment
([Rb/M]=1.0$\pm$0.4). A more trustful comparison could be made
with a similar sample, but for stars belonging to MCs
\citep{gh09}. In that case, distances are better known and
bolometric magnitudes can be derived. On average, stars are
Zr-poor (some of them showing some enhancement) and, thus, agree
with our models. Those objects are Rb-rich. However, as for the
galactic sample, a decrease in the Rb abundance is expected when
considering a circum-stellar component, as demonstrated by the LMC
star (IRAS 04498-6842; \citealt{zamo14}). The rubidium surface
enrichment of this star ([Rb/M]=1.5$\pm$0.7) is not in agreement
with our models, even when taking into account the large
observational errors. Its bolometric magnitude ($M_{\rm
bol}$=-7.72) may indicate a larger stellar mass with respect to
those presented in this paper. At the same time, however, it could
be the proof of the activation of HBB, which implies larger
surface luminosities with respect to those expected from the
core-luminosity relation (see e.g. \citealt{bs91}). In conclusion,
apart from one single object, our models do not disagree with the
discussed observational data. Another interesting sample is that
by \cite{mcsave}, who presented C, N, and O abundances in two
O-rich luminous Giant belonging to the Large magellanic Cloud (NGC
1866\#4 and HV2576). In particular, those authors found a strong
carbon depletion ($\sim$ 1 dex) coupled to a clear nitrogen
enhancement ($\sim$ 1 dex). They concluded that this is the proof
for the occurrence of ongoing HBB in the analyzed stars. However,
alternative theories could be explored. First, we remember that,
before entering the TP-AGB phase, FDU and SDU increase the surface
N abundance by a factor of 3. This is not enough to reproduce
observations. However, it has to be stressed that additional
physical phenomena may produce similar abundance patterns. An
illuminating case is represented by rotation. Models presented in
this paper do not take into account the effects of rotation. In
\cite{pi13}, we demonstrated that mixing induced by rotation may
significantly change the final surface theoretical distributions
of LMS-AGB models. Rotation may induce non canonical mixing also
in larger masses. In particular, during the Main Sequence phase
meridional circulations \citep{vz2,vz1} may work in the layers
between the inner border of the convective envelope and the upper
border of the receding H-burning convective core. In that case, a
mixing would develop in a region that previously experienced CN
cycling. Later, after the occurrence of FDU, the surface CN
abundances could result varied. We test the effects of rotation on
a 6 M\odos model with $Z=10^{-2}$ and different initial rotation
velocities. We find that models rotating on the Zero Age Main
Sequence (ZAMS) with v$^{\rm ini}_{\rm rot}\sim $100 km/s (thus
not so high for stars with this mass) already show a large C
depletion (-1 dex) and a strong nitrogen enhancement (+0.7 dex).
Note that similar conclusions have  already been derived by
\cite{georgy13}, even with a different formulation for the
transport of
angular momentum.\\
Other useful constraints to theoretical models come from the study
of the abundances derived in Planetary Nebulae (PNe), which lie
between the tip of the AGB and the WDs cooling sequence (for a
review see \citealt{pn}). During that evolutionary phase, the
strong mass loss practically peels the H-exhausted core. The star
evolves toward higher surface temperatures at almost constant
luminosity and, when the ionization of the lost gas begins, a PN
emerges. It has been shown (see \citealt{gh14} for an updated
study) that a consistent fraction of PNe are N-rich (the so-called
type I PNe). The most $^{14}$N-enriched PNe are not accessible by
our models. However, a considerable fraction of the sample can be
reproduced by taking into account the variations in CNO abundances
caused by the occurrence of FDU and/or SDU. Moreover, the effects
induced by rotation (see above) or the presence of a companion
\citep{dema09} could complicate the physical and chemical behavior
of PNe. Finally, let us stress that a precise determination of the
PNe masses is not an easy task (as always for galactic objects).
It could be that the most N-rich PNe are the remnant of the
evolution of massive AGBs (7-8-9 M\odo; the so-called super-AGBs),
whose
evolution is not explored in this paper.\\
Finally, let us stress that up to date no signature of HBB has
been found in pre-solar grains. In 2007, it has been claimed that
the composition of the peculiar spinel grain OC2 could be
attributed to the nucleosynthesis induced by HBB in a massive AGB
star \citep{lugaro07}. However, in order to reconcile theoretical
models and laboratory measurements, a modification of nuclear
cross sections was needed. Later, the same authors \citep{ilia08}
rejected such an hypothesis, identifying a low mass star
experiencing additional non convective mixing as the best
candidate to explain the isotopic signatures of that grain (see
also \citealt{palme13}). This does not necessarily imply that
pre-solar SiC grains carrying the signature of HBB do not exist,
but that they have not been discovered yet.

\acknowledgments

This work was supported by the Italian Grant PRIN-MIUR 2012
``Nucleosynthesis in AGB stars: an integrated approach'' project
(20128PCN59). We warmly thank Dr. Quintini for helping in the
development and maintenance of the FRUITY web interface. We thank
the anonymous referee for a careful reading of the text and for
valuable comments.

\bibliographystyle{aa}
\bibliography{cristallo}

\begin{deluxetable}{ccccccccccc}
\rotate \tablecolumns{11} \tablewidth{0pc} \tablecaption{Initial
helium and $\alpha$ enrichment for different initial masses and
metallicities. Bold labels refer to models added with this work.}
\startdata {Mass$\diagup$Z} & $2.4\times10^{-4}$ &
$7.2\times10^{-4}$ & $1.0\times10^{-3}$ & $2.0\times10^{-3}$ &
$3.0\times10^{-3}$ & $6.0\times10^{-3}$& $8.0\times10^{-3}$ &
$1.0\times10^{-2}$& $1.4\times10^{-2}$ & $2.0\times10^{-2}$ \\
{Mass$\diagup$[Fe/H]} & -2.15 & -1.67 & -1.15 & -0.85 & -0.67 &
-0.37& -0.24 & -0.15& 0.00 & 0.15 \\\hline
 1.3  & {\small (0.245;0.5)} & {\small (0.245;0.5)}& {\small (0.245;0)}& {\small \textbf{(0.252;0)}}& {\small (0.260;0)}& {\small (0.260;0)}& {\small (0.265;0)}& {\small (0.265;0)}& {\small (0.269;0)}& {\small (0.269;0})\\
 1.5  & {\small (0.245;0.5)} & {\small (0.245;0.5)}& {\small (0.245;0)}& {\small \textbf{(0.252;0)}}& {\small (0.260;0)}& {\small (0.260;0)}& {\small (0.265;0)}& {\small (0.265;0)}& {\small (0.269;0)}& {\small (0.269;0})\\
 2.0  & {\small (0.245;0.5)} & {\small (0.245;0.5)}& {\small (0.245;0)}& {\small \textbf{(0.252;0)}}& {\small (0.260;0)}& {\small (0.260;0)}& {\small (0.265;0)}& {\small (0.265;0)}& {\small (0.269;0)}& {\small (0.269;0})\\
 2.5  & {\small (0.245;0.5)} & {\small (0.245;0.5)}& {\small (0.245;0)}& {\small \textbf{(0.252;0)}}& {\small (0.260;0)}& {\small (0.260;0)}& {\small (0.265;0)}& {\small (0.265;0)}& {\small (0.269;0)}& {\small (0.269;0})\\
 3.0  & {\small (0.245;0.5)} & {\small (0.245;0.5)}& {\small (0.245;0)}& {\small \textbf{(0.252;0)}}& {\small (0.260;0)}& {\small (0.260;0)}& {\small (0.265;0)}& {\small (0.265;0)}& {\small (0.269;0)}& {\small (0.269;0})\\
 4.0  & {\small \textbf{(0.245;0.5)}} & {\small (0.245;0.5)}& {\small \textbf{(0.245;0)}}& {\small \textbf{(0.252;0)}}& {\small \textbf{(0.260;0)}}& {\small \textbf{(0.260;0)}}& {\small \textbf{(0.265;0)}}& {\small \textbf{(0.265;0)}}& {\small \textbf{(0.269;0)}}& {\small \textbf{(0.269;0)}}\\
\enddata \label{tabini}
\end{deluxetable}

\begin{deluxetable}{ccccccccccc}
\rotate \tablecolumns{11} \tablewidth{0pc} \tablecaption{Main
Sequence lifetimes (yr) for different initial masses and
metallicities. }  \startdata {Mass$\diagup$[Fe/H]} & -2.15 & -1.67
& -1.15 & -0.85 & -0.67 & -0.37& -0.24 & -0.15& 0.00 & 0.15
\\ \hline
 1.3          & {\small 2.38E+09  }& {\small2.38E+09  }& {\small2.45E+09  }& {\small2.50E+09 }& {\small2.53E+09 }& {\small2.95E+09 }& {\small3.15E+09 }& {\small3.49E+09 }& {\small3.75E+09 }& {\small4.76E+09 }\\
 1.5          & {\small 1.51E+09  }& {\small1.51E+09  }& {\small1.56E+09  }& {\small1.60E+09 }& {\small1.61E+09 }& {\small1.85E+09 }& {\small1.98E+09 }& {\small2.16E+09 }& {\small2.31E+09 }& {\small2.84E+09 }\\
 2.0          & {\small 6.25E+08  }& {\small6.51E+08  }& {\small6.80E+08  }& {\small7.12E+08 }& {\small7.12E+08 }& {\small8.10E+08 }& {\small8.44E+08 }& {\small9.13E+08 }& {\small9.70E+08 }& {\small1.16E+09 }\\
 2.5          & {\small 3.44E+08  }& {\small3.60E+08  }& {\small3.75E+08  }& {\small3.88E+08 }& {\small3.82E+08 }& {\small4.30E+08 }& {\small4.47E+08 }& {\small4.79E+08 }& {\small5.03E+08 }& {\small5.92E+08 }\\
 3.0          & {\small 2.20E+08  }& {\small2.28E+08  }& {\small2.35E+08  }& {\small2.42E+08 }& {\small2.38E+08 }& {\small2.64E+08 }& {\small2.71E+08 }& {\small2.88E+08 }& {\small2.99E+08 }& {\small3.46E+08 }\\
 4.0          & {\small 1.14E+08  }& {\small1.16E+08  }& {\small1.20E+08  }& {\small1.20E+08 }& {\small1.19E+08 }& {\small1.29E+08 }& {\small1.32E+08 }& {\small1.39E+08 }& {\small1.44E+08 }& {\small1.66E+08 }\\
 5.0          & {\small 7.11E+07  }& {\small7.17E+07  }& {\small7.30E+07  }& {\small7.23E+07 }& {\small7.11E+07 }& {\small7.53E+07 }& {\small7.65E+07 }& {\small7.97E+07 }& {\small8.19E+07 }& {\small9.10E+07 }\\
\enddata \label{tabms}
\end{deluxetable}

\begin{deluxetable}{ccccccccccc}
\rotate \tablecolumns{11} \tablewidth{0pc} \tablecaption{Maximum
size of the convective core developed during the MS (in solar
units). }  \startdata {Mass$\diagup$[Fe/H]} & -2.15 & -1.67 &
-1.15 & -0.85 & -0.67 & -0.37& -0.24 &
-0.15& 0.00 & 0.15 \\
\hline
 1.3          & {\small  0.147 }& {\small 0.130}& {\small  0.130}& {\small 0.111}& {\small 0.120}& {\small 0.155}& {\small 0.164}& {\small 0.155}& {\small 0.164}& {\small 0.155}\\
 1.5          & {\small  0.211 }& {\small 0.200}& {\small  0.189}& {\small 0.169}& {\small 0.169}& {\small 0.147}& {\small 0.141}& {\small 0.136}& {\small 0.131}& {\small 0.122}\\
 2.0          & {\small  0.386 }& {\small 0.376}& {\small  0.365}& {\small 0.346}& {\small 0.336}& {\small 0.327}& {\small 0.310}& {\small 0.301}& {\small 0.293}& {\small 0.270}\\
 2.5          & {\small  0.569 }& {\small 0.553}& {\small  0.538}& {\small 0.524}& {\small 0.509}& {\small 0.496}& {\small 0.483}& {\small 0.496}& {\small 0.469}& {\small 0.444}\\
 3.0          & {\small  0.783 }& {\small 0.761}& {\small  0.741}& {\small 0.721}& {\small 0.702}& {\small 0.664}& {\small 0.663}& {\small 0.646}& {\small 0.628}& {\small 0.594}\\
 4.0          & {\small  1.198 }& {\small 1.198}& {\small  1.133}& {\small 1.074}& {\small 1.045}& {\small 1.015}& {\small 0.987}& {\small 0.962}& {\small 0.962}& {\small 0.988}\\
 5.0          & {\small  1.606 }& {\small 1.606}& {\small  1.538}& {\small 1.457}& {\small 1.417}& {\small 1.342}& {\small 1.342}& {\small 1.342}& {\small 1.269}& {\small 1.342}\\
 6.0          & {\small  2.043 }& {\small 2.043}& {\small  1.928}& {\small 1.846}& {\small 1.797}& {\small 1.700}& {\small 1.700}& {\small 1.700}& {\small 1.610}& {\small 1.700}\\
 \enddata \label{tabcch}
\end{deluxetable}

\begin{deluxetable}{ccccccccccc}
\rotate \tablecolumns{11} \tablewidth{0pc} \tablecaption{The H-exhausted
core mass at the beginning of the He burning (in solar units). }
\startdata {Mass$\diagup$[Fe/H]} & -2.15 & -1.67 & -1.15 & -0.85 &
-0.67 & -0.37& -0.24 &
-0.15& 0.00 & 0.15 \\
\hline
 1.3  & {\small 0.485}& {\small 0.487}& {\small 0.488}& {\small 0.487}& {\small 0.485}& {\small 0.485}& {\small 0.490}& {\small 0.490}& {\small 0.481}& {\small 0.480}\\
 1.5  & {\small 0.473}& {\small 0.482}& {\small 0.484}& {\small 0.484}& {\small 0.484}& {\small 0.484}& {\small 0.483}& {\small 0.483}& {\small 0.481}& {\small 0.480}\\
 2.0  & {\small 0.404}& {\small 0.437}& {\small 0.452}& {\small 0.451}& {\small 0.451}& {\small 0.462}& {\small 0.463}& {\small 0.466}& {\small 0.465}& {\small 0.471}\\
 2.5  & {\small 0.347}& {\small 0.344}& {\small 0.343}& {\small 0.341}& {\small 0.324}& {\small 0.317}& {\small 0.319}& {\small 0.317}& {\small 0.315}& {\small 0.314}\\
 3.0  & {\small 0.405}& {\small 0.400}& {\small 0.397}& {\small 0.393}& {\small 0.383}& {\small 0.383}& {\small 0.382}& {\small 0.378}& {\small 0.375}& {\small 0.369}\\
 4.0  & {\small 0.533}& {\small 0.533}& {\small 0.525}& {\small 0.518}& {\small 0.518}& {\small 0.506}& {\small 0.504}& {\small 0.498}& {\small 0.494}& {\small 0.484}\\
 5.0  & {\small 0.675}& {\small 0.666}& {\small 0.675}& {\small 0.656}& {\small 0.656}& {\small 0.639}& {\small 0.635}& {\small 0.630}& {\small 0.626}& {\small 0.618}\\
 6.0  & {\small 0.822}& {\small 0.822}& {\small 0.822}& {\small 0.799}& {\small 0.799}& {\small 0.782}& {\small 0.777}& {\small 0.766}& {\small 0.766}& {\small 0.756}\\
 \enddata \label{tabinihe}
\end{deluxetable}

\begin{deluxetable}{ccccccccccc}
\rotate \tablecolumns{11} \tablewidth{0pc} \tablecaption{Central
helium burning lifetimes (yr) for different initial masses and
metallicities. } \startdata {Mass$\diagup$[Fe/H]} & -2.15 & -1.67
& -1.15 & -0.85 & -0.67 & -0.37& -0.24 &
-0.15& 0.00 & 0.15 \\
\hline
 1.3          & {\small8.72E+07 } & {\small8.95E+07  } & {\small9.06E+07}  & {\small9.39E+07} & {\small9.64E+07} & {\small1.01E+08 }& {\small9.16E+07 }& {\small9.23E+07 }& {\small1.10E+08 }& {\small1.17E+08 }\\
 1.5          & {\small9.16E+07 } & {\small9.02E+07  } & {\small9.08E+07}  & {\small9.35E+07} & {\small9.57E+07} & {\small1.00E+06 }& {\small1.03E+08 }& {\small1.05E+08 }& {\small1.09E+08 }& {\small1.16E+08 }\\
 2.0          & {\small1.13E+08 } & {\small1.04E+08  } & {\small1.02E+08}  & {\small1.11E+08} & {\small1.14E+08} & {\small1.14E+08 }& {\small1.16E+08 }& {\small1.17E+08 }& {\small1.19E+08 }& {\small1.22E+08 }\\
 2.5          & {\small9.68E+07 } & {\small1.11E+08  } & {\small1.23E+08}  & {\small1.36E+08} & {\small1.51E+08} & {\small1.99E+08 }& {\small2.18E+08 }& {\small2.42E+08 }& {\small2.59E+08 }& {\small3.02E+08 }\\
 3.0          & {\small5.87E+07 } & {\small6.45E+07  } & {\small7.05E+07}  & {\small7.70E+07} & {\small7.96E+07} & {\small9.75E+07 }& {\small1.10E+08 }& {\small1.25E+08 }& {\small1.38E+08 }& {\small1.68E+08 }\\
 4.0          & {\small2.67E+07 } & {\small2.81E+07  } & {\small3.02E+07}  & {\small3.28E+07} & {\small3.41E+07} & {\small3.79E+07 }& {\small3.92E+07 }& {\small4.21E+07 }& {\small4.54E+07 }& {\small5.68E+07 }\\
 5.0          & {\small1.50E+07 } & {\small1.56E+07  } & {\small1.66E+07}  & {\small1.80E+07} & {\small1.87E+07} & {\small2.06E+07 }& {\small2.09E+07 }& {\small2.19E+07 }& {\small2.21E+07 }& {\small2.47E+07 }\\
\enddata \label{tabheb}
\end{deluxetable}

\begin{deluxetable}{ccccccccccc}
\rotate \tablecolumns{11} \tablewidth{0pc} \tablecaption{The H-exhausted
core mass at the end of the He burning (in solar units). }
\startdata {Mass$\diagup$[Fe/H]} & -2.15 & -1.67 & -1.15 & -0.85 &
-0.67 & -0.37& -0.24 &
-0.15& 0.00 & 0.15 \\
\hline
 1.3  & {\small 0.551}& {\small 0.548}& {\small 0.544}& {\small 0.540}& {\small 0.538}& {\small 0.533}& {\small 0.531}& {\small 0.528}& {\small 0.527}& {\small 0.523}\\
 1.5  & {\small 0.567}& {\small 0.561}& {\small 0.553}& {\small 0.547}& {\small 0.545}& {\small 0.537}& {\small 0.535}& {\small 0.532}& {\small 0.531}& {\small 0.525}\\
 2.0  & {\small 0.616}& {\small 0.604}& {\small 0.588}& {\small 0.556}& {\small 0.550}& {\small 0.539}& {\small 0.534}& {\small 0.531}& {\small 0.531}& {\small 0.527}\\
 2.5  & {\small 0.716}& {\small 0.684}& {\small 0.660}& {\small 0.621}& {\small 0.595}& {\small 0.531}& {\small 0.517}& {\small 0.506}& {\small 0.506}& {\small 0.489}\\
 3.0  & {\small 0.819}& {\small 0.797}& {\small 0.775}& {\small 0.736}& {\small 0.717}& {\small 0.642}& {\small 0.607}& {\small 0.577}& {\small 0.562}& {\small 0.536}\\
 4.0  & {\small 1.055}& {\small 1.045}& {\small 1.026}& {\small 0.991}& {\small 0.968}& {\small 0.888}& {\small 0.862}& {\small 0.824}& {\small 0.793}& {\small 0.715}\\
 5.0  & {\small 1.324}& {\small 1.319}& {\small 1.300}& {\small 1.269}& {\small 1.247}& {\small 1.133}& {\small 1.075}& {\small 1.066}& {\small 1.052}& {\small 0.976}\\
 6.0  & {\small 1.610}& {\small 1.600}& {\small 1.578}& {\small 1.556}& {\small 1.534}& {\small 1.423}& {\small 1.355}& {\small 1.286}& {\small 1.325}& {\small 1.245}\\
 \enddata \label{tabheend}
\end{deluxetable}

\begin{deluxetable}{ccccccccccc}
\rotate \tablecolumns{11} \tablewidth{0pc} \tablecaption{ The total stellar mass at the beginning of the AGB phase (in solar units). }
\startdata {Mass$\diagup$[Fe/H]} & -2.15 & -1.67 & -1.15 & -0.85 &
-0.67 & -0.37& -0.24 &
-0.15& 0.00 & 0.15 \\
\hline
 1.3  & {\small  1.203}& {\small 1.187}& {\small 1.177}& {\small 1.165}& {\small 1.157}& {\small 1.134}& {\small 1.124}& {\small 1.117}& {\small 1.109}& {\small 1.091}\\
 1.5  & {\small  1.422}& {\small 1.405}& {\small 1.398}& {\small 1.388}& {\small 1.379}& {\small 1.362}& {\small 1.354}& {\small 1.350}& {\small 1.343}& {\small 1.330}\\
 2.0  & {\small  1.956}& {\small 1.942}& {\small 1.938}& {\small 1.941}& {\small 1.938}& {\small 1.928}& {\small 1.921}& {\small 1.915}& {\small 1.917}& {\small 1.899}\\
 2.5  & {\small  2.458}& {\small 2.456}& {\small 2.457}& {\small 2.461}& {\small 2.466}& {\small 2.477}& {\small 2.478}& {\small 2.479}& {\small 2.475}& {\small 2.473}\\
 3.0  & {\small  2.957}& {\small 2.949}& {\small 2.948}& {\small 2.950}& {\small 2.949}& {\small 2.963}& {\small 2.969}& {\small 2.976}& {\small 2.978}& {\small 2.981}\\
 4.0  & {\small  3.963}& {\small 3.943}& {\small 3.920}& {\small 3.894}& {\small 3.837}& {\small 3.785}& {\small 3.768}& {\small 3.807}& {\small 3.822}& {\small 3.940}\\
 5.0  & {\small  4.952}& {\small 4.909}& {\small 4.862}& {\small 4.800}& {\small 4.713}& {\small 4.578}& {\small 4.599}& {\small 4.556}& {\small 4.516}& {\small 4.517}\\
 6.0  & {\small  5.886}& {\small 5.797}& {\small 5.772}& {\small 5.645}& {\small 5.467}& {\small 5.203}& {\small 5.203}& {\small 5.266}& {\small 5.103}& {\small 4.970}\\
 \enddata \label{tabiniagb}
\end{deluxetable}

\begin{deluxetable}{ccccccccccc}
\rotate \tablecolumns{11} \tablewidth{0pc} \tablecaption{The H-exhausted
core mass at the beginning of the AGB phase (in solar units).}
\startdata {Mass$\diagup$[Fe/H]} & -2.15 & -1.67 & -1.15 & -0.85 &
-0.67 & -0.37& -0.24 &
-0.15& 0.00 & 0.15 \\
\hline
 1.3  & {\small 0.563}& {\small 0.559}& {\small 0.554}& {\small 0.551}& {\small 0.550}& {\small 0.543}& {\small 0.541}& {\small 0.538}& {\small 0.539}& {\small 0.534}\\
 1.5  & {\small 0.578}& {\small 0.572}& {\small 0.564}& {\small 0.557}& {\small 0.555}& {\small 0.547}& {\small 0.545}& {\small 0.543}& {\small 0.542}& {\small 0.536}\\
 2.0  & {\small 0.629}& {\small 0.615}& {\small 0.596}& {\small 0.566}& {\small 0.557}& {\small 0.545}& {\small 0.543}& {\small 0.541}& {\small 0.541}& {\small 0.538}\\
 2.5  & {\small 0.724}& {\small 0.697}& {\small 0.670}& {\small 0.633}& {\small 0.598}& {\small 0.540}& {\small 0.528}& {\small 0.519}& {\small 0.523}& {\small 0.517}\\
 3.0  & {\small 0.798}& {\small 0.795}& {\small 0.773}& {\small 0.742}& {\small 0.721}& {\small 0.651}& {\small 0.616}& {\small 0.585}& {\small 0.572}& {\small 0.545}\\
 4.0  & {\small 0.857}& {\small 0.857}& {\small 0.853}& {\small 0.845}& {\small 0.840}& {\small 0.817}& {\small 0.811}& {\small 0.796}& {\small 0.785}& {\small 0.722}\\
 5.0  & {\small 0.928}& {\small 0.927}& {\small 0.920}& {\small 0.912}& {\small 0.905}& {\small 0.874}& {\small 0.861}& {\small 0.857}& {\small 0.850}& {\small 0.826}\\
 6.0  & {\small 1.031}& {\small 1.028}& {\small 1.018}& {\small 1.006}& {\small 0.999}& {\small 0.959}& {\small 0.935}& {\small 0.913}& {\small 0.920}& {\small 0.893}\\
 \enddata \label{tabmtotiniagb}
\end{deluxetable}

\begin{deluxetable}{cccccccccc}
\rotate \tablecolumns{10} \tablewidth{0pc} \tablecaption{ Surface mass fractions of selected isotopes for models with $Z=1.0\times 10^{-2}$ at different evolutionary phases (initial, after the FDU, after the SDU). }
\startdata
\hline
Isot. & Phase &     1.3     &     1.5    &      2.0    &      2.5     &     3.0     &     4.0    &      5.0     &     6.0    \\
\hline
$^{4}$He     & INI   &  {\small  2.65E-01}   &  {\small  2.65E-01}  &   {\small  2.65E-01}  &   {\small  2.65E-01}   &  {\small  2.65E-01}   & {\small   2.65E-01}  &   {\small  2.65E-01}   &  {\small  2.65E-01}  \\
$^{4}$He     & FDU   &  {\small  2.88E-01}   &  {\small  2.84E-01}  &   {\small  2.77E-01}  &   {\small  2.79E-01}   &  {\small  2.81E-01}   & {\small   2.76E-01}  &   {\small  2.71E-01}   &  {\small  2.70E-01}  \\
$^{4}$He     & SDU   &  {\small     ---  }   &  {\small     ---  }  &   {\small     ---  }  &   {\small     ---  }   &  {\small     ---  }   & {\small   2.85E-01}  &   {\small  3.11E-01}   &  {\small  3.31E-01}  \\
\hline
$^{12}$C     & INI   &  {\small  1.64E-03}   &  {\small  1.64E-03}  &   {\small  1.64E-03}  &   {\small  1.64E-03}   &  {\small  1.64E-03}   & {\small   1.64E-03}  &   {\small  1.64E-03}   &  {\small  1.64E-03}  \\
$^{12}$C     & FDU   &  {\small  1.19E-03}   &  {\small  1.12E-03}  &   {\small  1.01E-03}  &   {\small  9.60E-04}   &  {\small  9.55E-04}   & {\small   9.72E-04}  &   {\small  9.87E-04}   &  {\small  9.92E-04}  \\
$^{12}$C     & SDU   &  {\small     ---  }   &  {\small     ---  }  &   {\small     ---  }  &   {\small     ---  }   &  {\small     ---  }   & {\small   9.34E-04}  &   {\small  9.13E-04}   &  {\small  8.93E-04}  \\
\hline
$^{13}$C     & INI   &  {\small  1.99E-05}   &  {\small  1.99E-05}  &   {\small  1.99E-05}  &   {\small  1.99E-05}   &  {\small  1.99E-05}   & {\small   1.99E-05}  &   {\small  1.99E-05}   &  {\small  1.99E-05}  \\
\enddata \label{tabfdusdu}
\end{deluxetable}

\begin{deluxetable}{cccccccccc}
\rotate \tablecolumns{10} \tablewidth{0pc} \tablecaption{ Surface mass fractions of selected isotopes for models with $Z=2.3\times 10^{-4}$ at different evolutionary phases (initial, after the FDU, after the SDU). }
\startdata
\hline
Isot.       &   Phase &         1.3  &                 1.5   &                2.0    &           2.5  &                    3.0    &           4.0    &             5.0    &               6.0  \\
\hline
$^{4}$He     &   INI   & {\small2.45E-01} &    {\small2.45E-01 } &   {\small2.45E-01 }  &  {\small2.45E-01} &   {\small 2.45E-01  } &  {\small2.45E-01  } &  {\small2.45E-01 }  &  {\small2.45E-01} \\
$^{4}$He     &   FDU   & {\small2.68E-01} &    {\small2.69E-01 } &   {\small2.66E-01 }  &  {\small2.46E-01} &   {\small    ---    } &  {\small   ---    } &  {\small   ---   }  &  {\small   ---  } \\
$^{4}$He     &   SDU   & {\small   ---  } &    {\small   ---   } &   {\small   ---   }  &  {\small2.55E-01} &   {\small 2.63E-01  } &  {\small3.03E-01  } &  {\small3.28E-01 }  &  {\small3.42E-01} \\
\hline
$^{12}$C     &   INI   & {\small1.68E-05} &    {\small1.68E-05 } &   {\small1.68E-05 }  &  {\small1.68E-05} &   {\small 1.68E-05  } &  {\small1.68E-05  } &  {\small1.68E-05 }  &  {\small1.68E-05} \\
$^{12}$C     &   FDU   & {\small1.13E-05} &    {\small1.02E-05 } &   {\small8.62E-06 }  &  {\small1.44E-05} &   {\small    ---    } &  {\small   ---    } &  {\small   ---   }  &  {\small   ---  } \\
$^{12}$C     &   SDU   & {\small   ---  } &    {\small   ---   } &   {\small   ---   }  &  {\small8.76E-06} &   {\small 8.03E-06  } &  {\small7.13E-06  } &  {\small6.71E-06 }  &  {\small1.06E-05} \\
\hline
$^{13}$C     &   INI   & {\small2.04E-07} &    {\small2.04E-07 } &   {\small2.04E-07 }  &  {\small2.04E-07} &   {\small 2.04E-07  } &  {\small2.04E-07 } &   {\small2.04E-07}   &  {\small2.04E-07} \\
\enddata \label{tabfdusdu2}
\end{deluxetable}

\begin{deluxetable}{|c|lll|lll|lll|lll|lll|lll|lll|lll|lll|lll|}
\rotate \tablecolumns{31} \tablewidth{0pc}
\tablecaption{Description of dredge up events: first dredge up
(F), second dredge up (S) and third dredge ups (T) characterizing
models with different masses and metallicities. If the model
experience third dredge ups, we report the numbers of experienced
episodes. } \startdata {[Fe/H]} & \multicolumn{3}{c}{-2.15} &
\multicolumn{3}{c}{-1.67} &
 \multicolumn{3}{c}{-1.15} & \multicolumn{3}{c}{-0.85} & \multicolumn{3}{c}{-0.67} & \multicolumn{3}{c}{-0.37}& \multicolumn{3}{c}{0.24} &
\multicolumn{3}{c}{-0.15}& \multicolumn{3}{c}{0.00} & \multicolumn{3}{c}{0.15} \\
\hline Mass &{\small F}&{\small S}&{\small T}&{\small F} & {\small
S} & {\small T} &{\small F} & {\small S} & {\small T} &{\small F}
& {\small S} & {\small T} &
 {\small F} & {\small S} & {\small T} &{\small F} & {\small S} & {\small T} &{\small F} & {\small S} & {\small T} &{\small F} & {\small S} & {\small T} &
 {\small F} & {\small S} & {\small T} &{\small F} & {\small S} & {\small T} \\
\hline   1.3   &{\tiny YES} & {\tiny NO} & {\tiny 9} &{\tiny YES} &
{\tiny NO} & {\tiny 7}
 & {\tiny YES} & {\tiny NO} & {\tiny 5} & {\tiny YES} & {\tiny NO} & {\tiny 5}
 &{\tiny YES} & {\tiny NO} & {\tiny 4} & {\tiny YES} & {\tiny NO} & {\tiny 3}
 & {\tiny YES} & {\tiny NO} & {\tiny 2} & {\tiny YES} & {\tiny NO} & {\tiny 2}
 &{\tiny YES} & {\tiny NO} & {\tiny NO} & {\tiny YES} & {\tiny NO} & {\tiny NO}
 \\
 1.5   &{\tiny YES} & {\tiny NO} & {\tiny 12} &{\tiny YES} & {\tiny NO} & {\tiny 9}
 & {\tiny YES} & {\tiny NO} & {\tiny 8} & {\tiny YES} & {\tiny NO} & {\tiny 7}
 &{\tiny YES} & {\tiny NO} & {\tiny 6} & {\tiny YES} & {\tiny NO} & {\tiny 7}
 & {\tiny YES} & {\tiny NO} & {\tiny 5} & {\tiny YES} & {\tiny NO} & {\tiny 4}
 &{\tiny YES} & {\tiny NO} & {\tiny 5} & {\tiny YES} & {\tiny NO} & {\tiny 4}
 \\
 2.0   &{\tiny YES} & {\tiny NO} & {\tiny 13} &{\tiny YES} & {\tiny NO} & {\tiny 13}
 & {\tiny YES} & {\tiny NO} & {\tiny 11} & {\tiny YES} & {\tiny NO} & {\tiny 11}
 &{\tiny YES} & {\tiny NO} & {\tiny 11} & {\tiny YES} & {\tiny NO} & {\tiny 10}
 & {\tiny YES} & {\tiny NO} & {\tiny 10} & {\tiny YES} & {\tiny NO} & {\tiny 9}
 &{\tiny YES} & {\tiny NO} & {\tiny 9} & {\tiny YES} & {\tiny NO} & {\tiny 9}
 \\
\enddata \label{tabtdu}
\end{deluxetable}

\begin{deluxetable}{ccccccccccc}
\rotate \tablecolumns{11} \tablewidth{0pc}
\tablecaption{Cumulative dredged up mass (in M$_\odot$) for
different initial masses and metallicities. } \startdata
{Mass$\diagup$[Fe/H]} & -2.15 & -1.67 & -1.15 & -0.85 & -0.67 &
-0.37& -0.24 &
-0.15& 0.00 & 0.15 \\
\hline
1.3   & {\small1.36E-02 }& {\small1.19E-02 }& {\small9.89E-03} & {\small8.64E-03 }& {\small4.65E-03 }& {\small2.82E-03 }& {\small5.75E-04 }& {\small2.64E-04 }& {\small0.00E+00 }& {\small0.00E+00  }\\
1.5   & {\small3.28E-02 }& {\small2.83E-02 }& {\small2.45E-02} & {\small2.14E-02 }& {\small1.64E-02 }& {\small1.41E-02 }& {\small8.15E-03 }& {\small8.46E-03 }& {\small6.66E-03 }& {\small5.58E-03  }\\
2.0   & {\small7.89E-02 }& {\small6.44E-02 }& {\small6.54E-02} & {\small6.15E-02 }& {\small5.26E-02 }& {\small4.61E-02 }& {\small3.50E-02 }& {\small3.88E-02 }& {\small2.91E-02 }& {\small2.96E-02  }\\
2.5   & {\small6.61E-02 }& {\small4.54E-02 }& {\small5.88E-02} & {\small6.12E-02 }& {\small6.40E-02 }& {\small8.20E-02 }& {\small7.23E-02 }& {\small7.25E-02 }& {\small5.91E-02 }& {\small6.67E-02  }\\
3.0   & {\small4.83E-02 }& {\small2.95E-02 }& {\small3.17E-02} & {\small3.20E-02 }& {\small2.72E-02 }& {\small4.10E-02 }& {\small4.74E-02 }& {\small5.77E-02 }& {\small6.01E-02 }& {\small6.43E-02  }\\
4.0   & {\small5.15E-02 }& {\small3.15E-02 }& {\small2.17E-02} & {\small2.27E-02 }& {\small1.59E-02 }& {\small1.53E-02 }& {\small8.90E-03 }& {\small8.59E-03 }& {\small7.28E-03 }& {\small1.10E-02  }\\
5.0   & {\small4.31E-02 }& {\small3.01E-02 }& {\small2.16E-02} & {\small2.15E-02 }& {\small1.52E-02 }& {\small9.38E-03 }& {\small9.80E-03 }& {\small7.62E-03 }& {\small5.71E-03 }& {\small4.06E-03  }\\
6.0   & {\small1.64E-02 }& {\small1.77E-02 }& {\small1.84E-02} & {\small1.26E-02 }& {\small9.32E-03 }& {\small8.36E-03 }& {\small8.61E-03 }& {\small8.47E-03 }& {\small5.52E-03 }& {\small3.65E-03  }\\
 \enddata \label{tab_dmtdu}
\end{deluxetable}

\begin{deluxetable}{|c|cc|cc|cc|cc|cc|cc|cc|cc|cc|cc|}
\rotate \tablecolumns{21} \tablewidth{0pc} \tablecaption{TP-AGB
lifetimes (in units of 10$^4$ yr) and percentage spent in the
C-rich regime for different initial masses and metallicities. }
\startdata {[Fe/H]} & \multicolumn{2}{c}{-2.15} &
\multicolumn{2}{c}{-1.67} &
 \multicolumn{2}{c}{-1.15} & \multicolumn{2}{c}{-0.85} & \multicolumn{2}{c}{-0.67} & \multicolumn{2}{c}{-0.37}& \multicolumn{2}{c}{0.24} &
\multicolumn{2}{c}{-0.15}& \multicolumn{2}{c}{0.00} & \multicolumn{2}{c}{0.15} \\
\hline Mass &{\small $\tau$} &{\small C(\%)}&{\small $\tau$} &
{\small C(\%)} &{\small $\tau$} & {\small C(\%)} &{\small $\tau$}
& {\small C(\%)} &
 {\small $\tau$} & {\small C(\%)} &{\small $\tau$} & {\small C(\%)} &{\small $\tau$} & {\small C(\%)} &{\small $\tau$}  & {\small C(\%)} &
 {\small $\tau$} & {\small C(\%)} &{\small $\tau$} & {\small C(\%)} \\
\hline
1.3 &{\small 176} &{\small 79}&{\small 191} & {\small 62} &{\small 180} & {\small 57} &{\small 181} & {\small 57} & {\small 175} & {\small 47} &{\small 166} & {\small 25} &{\small 170} & {\small 0} &{\small 170}  & {\small 0} & {\small 152} & {\small 0} &{\small 156} & {\small 0} \\
1.5 &{\small 163} &{\small 74}&{\small 171} & {\small 71} &{\small 160} & {\small 62} &{\small 182} & {\small 53} & {\small 173} & {\small 48} &{\small 175} & {\small 41} &{\small 181} & {\small 33} &{\small 150}  & {\small 10} & {\small 156} & {\small 0} &{\small 183} & {\small 0} \\
2.0 &{\small 129} &{\small 84}&{\small 125} & {\small 77} &{\small 134} & {\small 79} &{\small 202} & {\small 68} & {\small 213} & {\small 60} &{\small 226} & {\small 45} &{\small 219} & {\small 35} &{\small 219}  & {\small 38} & {\small 206} & {\small 24} &{\small 202} & {\small 11} \\
2.5 &{\small 64.0} &{\small 89}&{\small 69.1} & {\small 84} &{\small 70.2} & {\small 83} &{\small 108} & {\small 76} & {\small 147} & {\small 67} &{\small 283} & {\small 59} &{\small 313} & {\small 47} &{\small 335}  & {\small 44} & {\small 296} & {\small 34} &{\small 316} & {\small 26} \\
\enddata \label{tab_agb}
\end{deluxetable}

\begin{deluxetable}{ccccccccccc}
\rotate \tablecolumns{11} \tablewidth{0pc} \tablecaption{Final
surface [ls/Fe] (representative of the first \s-process peak) for
different initial masses and metallicities. } \startdata
{Mass$\diagup$[Fe/H]} & -2.15 & -1.67 & -1.15 & -0.85 & -0.67 &
-0.37& -0.24 &
-0.15& 0.00 & 0.15 \\
\hline
 1.3  & {\small9.82E-01  }& {\small8.22E-01 }&  {\small4.80E-01} &  {\small5.53E-01 }&  {\small4.26E-01 }&  {\small4.22E-01 }& {\small1.22E-01 }& {\small5.20E-02 }& {\small0.00E+00 }& {\small0.00E+00  }\\
 1.5  & {\small1.17E+00  }& {\small1.07E+00 }&  {\small7.06E-01} &  {\small8.05E-01 }&  {\small7.27E-01 }&  {\small8.27E-01 }& {\small6.95E-01 }& {\small7.23E-01 }& {\small6.27E-01 }& {\small4.39E-01  }\\
 2.0  & {\small1.05E+00  }& {\small9.51E-01 }&  {\small1.03E+00} &  {\small1.03E+00 }&  {\small1.06E+00 }&  {\small1.17E+00 }& {\small1.12E+00 }& {\small1.12E+00 }& {\small1.02E+00 }& {\small9.28E-01  }\\
 2.5  & {\small1.23E+00  }& {\small7.39E-01 }&  {\small9.15E-01} &  {\small9.03E-01 }&  {\small1.02E+00 }&  {\small1.28E+00 }& {\small1.31E+00 }& {\small1.29E+00 }& {\small1.19E+00 }& {\small1.08E+00  }\\
 3.0  & {\small1.32E+00  }& {\small1.03E+00 }&  {\small8.53E-01} &  {\small8.11E-01 }&  {\small6.79E-01 }&  {\small8.45E-01 }& {\small1.03E+00 }& {\small1.05E+00 }& {\small1.06E+00 }& {\small1.01E+00  }\\
 4.0  & {\small1.48E+00  }& {\small8.53E-01 }&  {\small5.88E-01} &  {\small4.41E-01 }&  {\small4.82E-01 }&  {\small7.00E-01 }& {\small3.96E-01 }& {\small4.30E-01 }& {\small4.16E-01 }& {\small5.66E-01  }\\
 5.0  & {\small1.47E+00  }& {\small9.02E-01 }&  {\small4.95E-01} &  {\small3.28E-01 }&  {\small2.43E-01 }&  {\small2.14E-01 }& {\small2.58E-01 }& {\small2.07E-01 }& {\small1.62E-01 }& {\small1.22E-01  }\\
\enddata \label{tab_ls}
\end{deluxetable}
\begin{deluxetable}{ccccccccccc}
\rotate \tablecolumns{11} \tablewidth{0pc} \tablecaption{Final
surface [hs/Fe] (representative of the second \s-process peak) for
different initial masses and metallicities. } \startdata
{Mass$\diagup$[Fe/H]} & -2.15 & -1.67 & -1.15 & -0.85 & -0.67 &
-0.37& -0.24 &
-0.15& 0.00 & 0.15 \\
\hline
 1.3  & {\small1.19E+00 } & {\small1.11E+00}  & {\small1.00E+00 }& {\small1.14E+00  }& {\small9.62E-01  }& {\small 7.16E-01} &  {\small1.65E-01} &  {\small2.50E-02 }&{\small 0.00E+00 }&{\small 0.00E+00 } \\
 1.5  & {\small1.44E+00 } & {\small1.41E+00}  & {\small1.29E+00 }& {\small1.46E+00  }& {\small1.41E+00  }& {\small 1.31E+00} &  {\small9.71E-01} &  {\small7.83E-01 }&{\small 4.54E-01 }&{\small 1.15E-01 } \\
 2.0  & {\small1.54E+00 } & {\small1.47E+00}  & {\small1.57E+00 }& {\small1.67E+00  }& {\small1.75E+00  }& {\small 1.60E+00} &  {\small1.37E+00} &  {\small1.19E+00 }&{\small 7.94E-01 }&{\small 4.50E-01 } \\
 2.5  & {\small1.56E+00 } & {\small1.20E+00}  & {\small1.43E+00 }& {\small1.48E+00  }& {\small1.66E+00  }& {\small 1.69E+00} &  {\small1.50E+00} &  {\small1.27E+00 }&{\small 8.87E-01 }&{\small 5.42E-01 } \\
 3.0  & {\small1.41E+00 } & {\small1.39E+00}  & {\small1.29E+00 }& {\small1.27E+00  }& {\small1.27E+00  }& {\small 1.36E+00} &  {\small1.28E+00} &  {\small1.08E+00 }&{\small 8.07E-01 }&{\small 5.03E-01 } \\
 4.0  & {\small1.26E+00 } & {\small9.01E-01}  & {\small7.94E-01 }& {\small7.81E-01  }& {\small7.84E-01  }& {\small 8.38E-01} &  {\small6.68E-01} &  {\small5.99E-01 }&{\small 3.92E-01 }&{\small 3.55E-01 } \\
 5.0  & {\small1.22E+00 } & {\small5.49E-01}  & {\small3.51E-01 }& {\small3.90E-01  }& {\small4.26E-01  }& {\small 4.63E-01} &  {\small4.92E-01} &  {\small3.51E-01 }&{\small 1.76E-01 }&{\small 4.80E-02 } \\
\enddata \label{tab_hs}
\end{deluxetable}
\begin{deluxetable}{ccccccccccc}
\rotate \tablecolumns{11} \tablewidth{0pc} \tablecaption{Final
surface [Pb/Fe] (representative of the third \s-process peak)  for
different initial masses and metallicities. } \startdata
{Mass$\diagup$[Fe/H]} & -2.15 & -1.67 & -1.15 & -0.85 & -0.67 &
-0.37& -0.24 &
-0.15& 0.00 & 0.15 \\
\hline
 1.3  & {\small2.70E+00}  & {\small2.52E+00}  & {\small2.21E+00} &  {\small2.08E+00} &  {\small1.49E+00 }&  {\small5.88E-01 }&  {\small3.90E-02 }&  {\small3.00E-03 }&  {\small0.00E+00 }&  {\small0.00E+00 }\\
 1.5  & {\small2.88E+00}  & {\small2.80E+00}  & {\small2.46E+00} &  {\small2.38E+00} &  {\small1.93E+00 }&  {\small1.17E+00 }&  {\small6.42E-01 }&  {\small3.99E-01 }&  {\small1.52E-01 }&  {\small2.00E-02 }\\
 2.0  & {\small2.95E+00}  & {\small2.85E+00}  & {\small2.65E+00} &  {\small2.55E+00} &  {\small2.19E+00 }&  {\small1.47E+00 }&  {\small1.11E+00 }&  {\small8.57E-01 }&  {\small4.96E-01 }&  {\small1.89E-01 }\\
 2.5  & {\small2.91E+00}  & {\small2.66E+00}  & {\small2.53E+00} &  {\small2.45E+00} &  {\small2.13E+00 }&  {\small1.57E+00 }&  {\small1.26E+00 }&  {\small1.00E+00 }&  {\small6.21E-01 }&  {\small2.56E-01 }\\
 3.0  & {\small2.67E+00}  & {\small2.57E+00}  & {\small2.39E+00} &  {\small2.29E+00} &  {\small2.03E+00 }&  {\small1.29E+00 }&  {\small1.04E+00 }&  {\small7.96E-01 }&  {\small5.39E-01 }&  {\small2.35E-01 }\\
 4.0  & {\small2.19E+00}  & {\small2.05E+00}  & {\small1.83E+00} &  {\small1.75E+00} &  {\small1.59E+00 }&  {\small1.04E+00 }&  {\small6.63E-01 }&  {\small4.68E-01 }&  {\small2.31E-01 }&  {\small1.89E-01 }\\
 5.0  & {\small1.66E+00}  & {\small1.37E+00}  & {\small1.23E+00} &  {\small1.24E+00} &  {\small1.18E+00 }&  {\small7.39E-01 }&  {\small5.22E-01 }&  {\small2.42E-01 }&  {\small6.10E-02 }&  {\small1.20E-02 }\\
\enddata \label{tab_pb}
\end{deluxetable}

\begin{deluxetable}{ccccccccccc}
\rotate \tablecolumns{11} \tablewidth{0pc}
\tablecaption{\s-process index [hs/ls] for different initial
masses and metallicities. } \startdata {Mass$\diagup$[Fe/H]} &
-2.15 & -1.67 & -1.15 & -0.85 & -0.67 & -0.37& -0.24 &
-0.15& 0.00 & 0.15 \\
\hline
 1.3  &{\small  2.10E-01 } &{\small 2.92E-01  }&{\small 5.20E-01} &{\small  5.81E-01} &{\small  5.37E-01} &{\small  2.94E-01 }&{\small  4.30E-02} &{\small -2.70E-02} &{\small  0.00E+00} &{\small  0.00E+00} \\
 1.5  &{\small  2.78E-01 } &{\small 3.40E-01  }&{\small 5.84E-01} &{\small  6.50E-01} &{\small  6.83E-01} &{\small  4.84E-01 }&{\small  2.76E-01} &{\small  6.00E-02} &{\small -1.73E-01} &{\small -3.24E-01} \\
 2.0  &{\small  4.87E-01 } &{\small 5.21E-01  }&{\small 5.38E-01} &{\small  6.34E-01} &{\small  6.95E-01} &{\small  4.37E-01 }&{\small  2.53E-01} &{\small  6.60E-02} &{\small -2.25E-01} &{\small -4.78E-01} \\
 2.5  &{\small  3.33E-01 } &{\small 4.56E-01  }&{\small 5.12E-01} &{\small  5.74E-01} &{\small  6.37E-01} &{\small  4.10E-01 }&{\small  1.89E-01} &{\small -2.00E-02} &{\small -3.01E-01} &{\small -5.38E-01} \\
 3.0  &{\small  8.80E-02 } &{\small 3.57E-01  }&{\small 4.40E-01} &{\small  4.59E-01} &{\small  5.92E-01} &{\small  5.12E-01 }&{\small  2.50E-01} &{\small  3.00E-02} &{\small -2.55E-01} &{\small -5.04E-01} \\
 4.0  &{\small -2.27E-01 } &{\small 4.80E-02  }&{\small 2.06E-01} &{\small  3.40E-01} &{\small  3.01E-01} &{\small  1.38E-01 }&{\small  2.72E-01} &{\small  1.70E-01} &{\small -2.40E-02} &{\small -2.11E-01} \\
 5.0  &{\small -2.48E-01 } &{\small-3.52E-01  }&{\small-1.44E-01} &{\small  6.20E-02} &{\small  1.83E-01} &{\small  2.49E-01 }&{\small  2.34E-01} &{\small  1.44E-01} &{\small  1.30E-02} &{\small -7.40E-02} \\
 6.0  &{\small -2.78E-01 } &{\small-3.86E-01  }&{\small-3.56E-01} &{\small -1.44E-01} &{\small -5.00E-03} &{\small  1.05E-01 }&{\small  1.31E-01} &{\small  1.12E-01} &{\small  1.30E-02} &{\small -3.00E-02} \\
 \enddata \label{tab_hsls}
\end{deluxetable}
\begin{deluxetable}{ccccccccccc}
\rotate \tablecolumns{11} \tablewidth{0pc}
\tablecaption{\s-process index [Pb/hs] for different initial
masses and metallicities. } \startdata {Mass$\diagup$[Fe/H]} &
-2.15 & -1.67 & -1.15 & -0.85 & -0.67 & -0.37& -0.24 &
-0.15& 0.00 & 0.15 \\
\hline
 1.3 &{\small 1.51E+00 }  &{\small 1.40E+00  }&{\small 1.21E+00 }&{\small  9.40E-01} &{\small  5.26E-01 }&{\small -1.28E-01 }&{\small -1.26E-01} &{\small -2.20E-02} &{\small  0.00E+00 }&{\small  0.00E+00 }\\
 1.5 &{\small 1.44E+00 }  &{\small 1.39E+00  }&{\small 1.17E+00 }&{\small  9.27E-01} &{\small  5.19E-01 }&{\small -1.45E-01 }&{\small -3.30E-01} &{\small -3.84E-01} &{\small -3.02E-01 }&{\small -9.50E-02 }\\
 2.0 &{\small 1.41E+00 }  &{\small 1.38E+00  }&{\small 1.08E+00 }&{\small  8.79E-01} &{\small  4.38E-01 }&{\small -1.35E-01 }&{\small -2.62E-01} &{\small -3.31E-01} &{\small -2.98E-01 }&{\small -2.61E-01 }\\
 2.5 &{\small 1.34E+00 }  &{\small 1.47E+00  }&{\small 1.10E+00 }&{\small  9.75E-01} &{\small  4.72E-01 }&{\small -1.21E-01 }&{\small -2.31E-01} &{\small -2.68E-01} &{\small -2.66E-01 }&{\small -2.86E-01 }\\
 3.0 &{\small 1.26E+00 }  &{\small 1.18E+00  }&{\small 1.09E+00 }&{\small  1.02E+00} &{\small  7.56E-01 }&{\small -7.10E-02 }&{\small -2.33E-01} &{\small -2.82E-01} &{\small -2.68E-01 }&{\small -2.68E-01 }\\
 4.0 &{\small 9.31E-01 }  &{\small 1.15E+00  }&{\small 1.04E+00 }&{\small  9.74E-01} &{\small  8.04E-01 }&{\small  1.98E-01 }&{\small -5.00E-03} &{\small -1.32E-01} &{\small -1.60E-01 }&{\small -1.66E-01 }\\
 5.0 &{\small 4.32E-01 }  &{\small 8.22E-01  }&{\small 8.76E-01 }&{\small  8.48E-01} &{\small  7.50E-01 }&{\small  2.76E-01 }&{\small  3.00E-02} &{\small -1.09E-01} &{\small -1.15E-01 }&{\small -3.60E-02 }\\
 6.0 &{\small 1.19E-01 }  &{\small 1.78E-01  }&{\small 2.84E-01 }&{\small  4.13E-01} &{\small  3.39E-01 }&{\small  2.16E-01 }&{\small  5.10E-02} &{\small -6.20E-02} &{\small -6.00E-02 }&{\small -1.50E-02 }\\
 \enddata \label{tab_pbhs}
\end{deluxetable}

\begin{deluxetable}{lccc}
\tablecolumns{4} \tablewidth{0pc} \tablecaption{Physical and chemical properties of our 5 M\odos model with $Z=10^{-3}$ compared to \citealt{fi14} (FI14) and
\citealt{veda08} (VD08).} \startdata
 & This paper&  FI14 & VD08 \\ \hline
M$^{\rm INI}_{\rm H}$ [M\odo]      & 0.92 & 0.92 & 0.91 \\
M$^{\rm END}_{\rm H}$ [M\odo]     & 0.94 & 0.95 & 0.94 \\
TP              & 25 & 94 & 53 \\
$\lambda_{\rm max}$ & 0.69 & 0.95 & 0.50 \\
$T^{\rm max}_{\rm CE}$  [K]  & $3.0\times 10^7$ &  $9.2\times 10^7$ & $1.0\times 10^8$ \\
Y              & 0.33 & 0.36 & 0.32 \\
$[{\rm C}/{\rm Fe}]$          & 1.16 & 0.36 & 0.13 \\
$[{\rm N}/{\rm Fe}]$         & 0.59 & 2.45 & 1.70 \\
$[{\rm O}/{\rm Fe}]$         & -0.05 & -0.20 & -0.06 \\
$[{\rm F}/{\rm Fe}]$         & -0.05 & 0.36 & -1.16 \\
$[{\rm Na}/{\rm Fe}]$        & 0.51 & 0.87 & 0.60 \\
$[{\rm Mg}/{\rm Fe}]$        & 0.07 & 0.61 & 0.35 \\
$[{\rm Al}/{\rm Fe}]$        & 0.04 & 0.74 & 1.02 \\
$[{\rm ls}/{\rm Fe}]$        & 0.50 & 1.33 & --- \\
$[{\rm hs}/{\rm Fe}]$        & 0.35 & 0.51 & --- \\
$[{\rm Pb}/{\rm Fe}]$        & 1.23 & 0.25 & --- \\
\enddata\label{tab_compa}
\end{deluxetable}

\begin{deluxetable}{ccccccccccc}
\rotate \tablecolumns{11} \tablewidth{0pc} \tablecaption{The H-exhausted core mass at
the end of the AGB phase (in solar units).} \startdata {Mass$\diagup$[Fe/H]} & -2.15 & -1.67
& -1.15 & -0.85 & -0.67 & -0.37& -0.24 &
-0.15& 0.00 & 0.15 \\
\hline
 1.3  &  {\small 0.658 } & {\small 0.631  }& {\small 0.628 }  &{\small 0.625  } &{\small 0.624  } &{\small 0.605 } & {\small 0.610} &  {\small 0.604}&   {\small 0.601} &  {\small 0.596}  \\
 1.5  &  {\small 0.667 } & {\small 0.645  }& {\small 0.635 }  &{\small 0.628  } &{\small 0.623  } &{\small 0.612 } & {\small 0.623} &  {\small 0.600}&   {\small 0.604} &  {\small 0.613}  \\
 2.0  &  {\small 0.676 } & {\small 0.671  }& {\small 0.656 }  &{\small 0.636  } &{\small 0.631  } &{\small 0.621 } & {\small 0.623} &  {\small 0.617}&   {\small 0.617} &  {\small 0.612}  \\
 2.5  &  {\small 0.749 } & {\small 0.724  }& {\small 0.699 }  &{\small 0.674  } &{\small 0.654  } &{\small 0.629 } & {\small 0.628} &  {\small 0.621}&   {\small 0.624} &  {\small 0.619}  \\
 3.0  &  {\small 0.815 } & {\small 0.810  }& {\small 0.789 }  &{\small 0.760  } &{\small 0.743  } &{\small 0.689 } & {\small 0.668} &  {\small 0.652}&   {\small 0.652} &  {\small 0.634}  \\
 4.0  &  {\small 0.876 } & {\small 0.874  }& {\small 0.867 }  &{\small 0.861  } &{\small 0.854  } &{\small 0.828 } & {\small 0.826} &  {\small 0.813}&   {\small 0.800} &  {\small 0.747}  \\
 5.0  &  {\small 0.949 } & {\small 0.946  }& {\small 0.936 }  &{\small 0.926  } &{\small 0.920  } &{\small 0.887 } & {\small 0.873} &  {\small 0.872}&   {\small 0.864} &  {\small 0.842}  \\
 6.0  &  {\small 1.060 } & {\small 1.050  }& {\small 1.040 }  &{\small 1.020  } &{\small 1.010  } &{\small 0.974 } & {\small 0.950} &  {\small 0.927}&   {\small 0.935} &  {\small 0.908}  \\
 \enddata \label{tab_mcore}
\end{deluxetable}

\end{document}